\documentclass[preprint,showpacs,preprintnumbers,amsmath,amssymb]{revtex4}

\usepackage{graphicx}
\usepackage{dcolumn}
\usepackage{bm}
\usepackage{subfigure}

\begin{document}

\title{ A FAST HADRON FREEZE-OUT GENERATOR }
\author{N.S. Amelin}
\altaffiliation[Also at ]{S. P. Korolev, Samara State Aerospace
University, Samara, 443086,  Russia}
\author{R. Lednicky}
\altaffiliation[Also at ]{Institute of Physics ASCR, Prague, 18221,
Czech Republic}
\author{T.A. Pocheptsov}
\affiliation{Joint Institute for Nuclear Research, Dubna, Moscow Region, 141980, Russia}
\author{ I.P. Lokhtin}
\author{L.V. Malinina}
\altaffiliation[Also at ]{Joint Institute for Nuclear Research, Dubna, Moscow Region, 141980, Russia}
\author{A.M. Snigirev}
\affiliation{M.V. Lomonosov Moscow State University, D.V. Skobeltsyn
Institute of Nuclear Physics, 119992, Moscow, Russia }
\author{Iu.A.~Karpenko}
\author{Yu.M. Sinyukov}
\affiliation{Bogolyubov Institute for Theoretical Physics, Kiev, 03143, Ukraine}

\date{\today}

\begin{abstract}
We have developed a fast Monte Carlo procedure of  hadron generation allowing one
to study and analyze various observables for stable hadrons and
hadron resonances produced in ultra-relativistic heavy ion
collisions.  Particle multiplicities are determined based on the concept of chemical freeze-out.
Particles can be generated on the chemical or thermal freeze-out hypersurface
represented by a parameterization or a numerical
solution of relativistic hydrodynamics with
given initial conditions and equation of state.
Besides standard space-like sectors associated with the volume decay,
the hypersurface may also include non-space-like sectors related
to the emission from the surface of expanding system.
For comparison with other models and experimental data
we demonstrate the results based on the standard
parameterizations of the hadron freeze-out hypersurface
and flow velocity profile under the assumption of a common chemical and
thermal freeze-out.
The C++ generator code is written under the ROOT
framework and is available for public use
at http://uhkm.jinr.ru/.
\end{abstract}



\maketitle
\section{\label{sec1}Introduction}

Ongoing and planned experimental studies of relativistic heavy ion
collisions in a wide range of beam energies require a development of new event generators and improvement
of the existing ones \cite{Heavy_Ions}.
Especially for Large Hadron Collider (LHC)
experiments,  because of very high hadron multiplicities, one needs fairly fast Monte-Carlo (MC) generators
for event simulation.

A successful but oversimplified attempt of creating a fast hadron
generator motivated by hydrodynamics was done  in
Ref.~\cite{Snigiriev96, Kruglov97, Lokhtin03, Lokhtin05}. The
present work is an extension of this approach. We formulate a fast
MC procedure to generate hadron multiplicities, four-momenta and
four-coordinates for any kind of freeze-out hypersurface. Decays of
hadronic resonances are taken into account. We consider hadrons
consisting of light $u$,$d$ and $s$ quarks only, but the extension
to heavier quarks is possible. The generator code is written in the
object-oriented C++ language under the ROOT framework \cite{ROOT}.

In this article we discuss only central collisions of nuclei using
the Bjorken-like and Hubble-like freeze-out parameterizations used
in so-called  "blast wave" \cite{Lisa} and "Cracow" models
\cite{Bron04}, respectively. The same parametrizations have been
used in the hadron generator referred as THERMINATOR
\cite{THERMINATOR} that appears however less efficient than our
generator (see sections \ref{sec2}, \ref{sec7}).

The paper is now organized as follows. Sections ~\ref{sec2}-\ref{sec6} are devoted to the
description of the physical framework of the model. In
section~\ref{sec7},  the
Monte Carlo simulation procedure is formulated. The validation of this
procedure is presented in section~\ref{sec8}. In section~\ref{sec9},  the example
calculations are compared  with the Relativistic Heavy Ion Collider
(RHIC) experimental
data. We summarize and conclude in section~\ref{sec10}.

\maketitle
\section{\label{sec2}Hadron multiplicities}

We give here the basic formulae for the calculation of
particle multiplicities. We consider the
hadronic matter created in heavy-ion collisions as a
hydrodynamically expanding fireball with the equation of state of an
ideal hadron gas.

The mean number $\bar{N}_i$ of particles species $i$ crossing the
\textit{space-like} freeze-out hypersurface $\sigma(x)$ in
Minkowski space can be computed as \cite{RKT}
\begin{equation}
\label{M1}
 \bar{N}_i =\int_{\sigma(x)} d^3\sigma_{\mu}(x) j^{\mu}_{i}(x).
\end{equation}
Here the four-vector $d^3\sigma_{\mu}(x)=n_{\mu}(x)d^3\sigma(x)$
is the element of  the freeze-out hypersurface
directed along the hypersurface normal unit
four-vector $n^{\mu}(x)$ with a positively defined zero component
($n^0(x)>0$) and $d^3\sigma(x)=|d^3\sigma_{\mu}d^3\sigma^{\mu}|^{1/2}$ is
the invariant  measure of this element.
The normal to the space-like hypersurface is time-like, i.e.
$n^{\mu}n_{\mu}=1$; generally, for hypersurfaces including
non-space-like sectors, the normal can also be a space-like so then
$n^{\mu}n_{\mu}=-1$.
The four-vector $j^{\mu}_{i}(x)$ is
the current of particle species $i$ determined as: \begin{equation}
\label{M2} j^{\mu}_{i}(x) = \int \frac{d^3\vec{ p}}{p^0} p^{\mu}
f_i(x,p),
\end{equation}
where $f_i(x,p)$  is the Lorentz invariant distribution function of
particle freeze-out four-coordinate $x = \{x^0,\vec{x}\}$ and
four-momentum $p = \{p^0,\vec{p}\}$.
In the case of local equilibrium
\begin{equation}
 \label{M2P}
 f_i(x,p)=f_i^{{\rm eq}}(p \cdot u(x);T(x),\mu(x))
=\frac{1}{(2\pi)^3}\frac{g_i}{\exp{([p \cdot u(x) - \mu_i(x)]/T(x))}
\pm 1},
\end{equation}
where $p \cdot u \equiv p^{\mu}u_{\mu}$,
$g_i=2 J_i+1$
is the spin degeneracy factor,
$T(x)$ and $\mu_{i}(x)$ are the local temperature
and chemical potential respectively,
$u(x)=\gamma\{1,\vec{v}\}$
is the local collective four-velocity, $\gamma = (1 - v^2)^{-1/2}$,
$u^{\mu}u_{\mu} = 1$. The signs $\pm$ in the denominator account for the
proper quantum statistics of a fermion or a boson, respectively.

The Lorentz scalar local particle density is defined as:
\begin{equation}
\label{M5}\rho_{i}(x)= u_{\mu}(x)j^{\mu}_{i}(x) = \int \frac{d^3\vec{p}}{p^0}
p_{\mu} u^{\mu}(x) f_i(x,p).
\end{equation}
For a system in local thermal equilibrium,
the particle density in the fluid element rest frame, where $u^{*\mu} =
\{1,0,0,0\}$, is solely determined by the local temperature
$T(x^{*})$ and chemical potential $\mu_i(x^{*})$ for each particle species
$i$:
\begin{equation}
\label{M6}
\rho_{i}^{\rm eq}(T(x^{*}),\mu_i(x^{*})) = u_{\mu}^{*}j^{{\rm eq}\mu}_{i}(x^*)
= \int d^3\vec{p}^{~*}
f_{i}^{\rm eq}(p^{*0}; T(x^{*}),\mu_i(x^{*}));
\end{equation}
the four-vectors in fluid element
rest frames are denoted by star.

In the case of local equilibrium, the particle current is proportional to the fluid
element four-velocity: $j^{{\rm eq} \mu}_{i}(x) = \rho_{i}^{\rm
eq}(T(x),\mu_i(x))u^{\mu}(x)$. So the mean number of particles of species $i$
is expressed directly through the equilibrated density:
\begin{equation}
\label{M7} \bar{N}_i =\int_{\sigma(x)} d^3\sigma_{\mu}(x)
u^{\mu}(x)\rho_i^{\rm eq}(T(x),\mu_i(x)).
\end{equation}

In the case of constant temperature and chemical potential,
$T(x) = T$ and $\mu_i(x) = \mu_i$,  one has
\begin{equation}
\label{M8} \bar{N}_i = \rho_{i}^{\rm eq}(T,\mu_i) \int_{\sigma(x)} d^3\sigma_{\mu}(x)
u^{\mu}(x) = \rho_{i}^{\rm eq}(T,\mu_i) V_{\rm eff},
\end{equation}
i.e. the total yield of particle species $i$ is determined by the
freeze-out temperature $T$, chemical potential $\mu_i$ and by the
total co-moving volume $V_{\rm eff}$, so called {\it effective
volume} of particle production which is a functional of the field of
collective velocities $u^{\mu}(x)$ on the hypersurface $\sigma (x)$.
The effective volume absorbs the collective velocity profile and the
form of hypersurface and cancels out in all particle number ratios.
Therefore, the particle number ratios do not depend on the
freeze-out details as long as the local thermodynamic parameters are
independent  of $x$. The concept of the effective volume and
factorization property similar to Eq.~(\ref{M8}) has been considered
first in Ref. \cite{Nukleonika}, repeatedly used for the analysis of
particle number ratios (see, e.g., Ref. \cite{Sinyukov02}) and
recently generalized for a study of the averaged phase space
densities \cite{AkkSin1} and entropy \cite{AkkSin2}. One can apply
this concept also in a limited rapidity window \cite{Nukleonika,
AkkSin1, AkkSin2}.

The concept of the effective volume can be applied to calculate the
hadronic composition at both chemical and thermal freeze-outs
\cite{Sinyukov02}. At the former one, which happens soon after
hadronization, the chemically equilibrated hadronic composition is
assumed to be established and frozen in further evolution. The
chemical potential $\mu_i$ for any particle species $i$
at the chemical freeze-out is entirely
determined by chemical potentials  $\widetilde{\mu_{q}}$ per a unit
charge, i.e. per unit baryon number $B$, strangeness $S$,
electric charge $Q$, charm $C$, etc. It can be expressed as a scalar
product:
\begin{equation}
\label{mui}
\mu_i =\vec{q}_i\vec{\widetilde{\mu}},
\end{equation}
where
$\vec{q}_i=\{ B_i, S_i, Q_i, C_i, ...\}$ and $\vec{\widetilde{\mu}}
=
 \{\widetilde{\mu}_{B}, \widetilde{\mu}_{S}, \widetilde{\mu}_{Q}, \widetilde{\mu}_{C},..\}$.
 Assuming constant temperature and
chemical potentials on the chemical freeze-out hypersurface,
the total quantum numbers $\vec{q}=\{B, S, Q,C,...\}$
of the selected thermal part of produced
hadronic system (e.g., in a rapidity interval near $y=0$) with
corresponding $V_{\rm eff}$ can be calculated as $\vec{q}=V_{\rm eff}
\sum_{i}\rho_{i}^{\rm eq}\vec{q}_i$.
For example:
\begin{equation}
B=\label{D16A} V_{\rm eff}\sum_{i=1}^{n}\rho_{i}^{\rm eq}(T,\mu_i) B_i,
\end{equation}

\begin{equation}
S=\label{D16} V_{\rm eff}\sum_{i=1}^{n} \rho_{i}^{\rm eq}(T,\mu_i) S_i,
\end{equation}

\begin{equation}
Q=\label{D17} V_{\rm eff}\sum_{i=1}^{n}\rho_{i}^{\rm eq}(T,\mu_i) Q_i.
\end{equation}
The potentials $\widetilde{\mu}_{q}$ are not independent. Thus,
taking into account baryon, strangeness and electrical charges only
and fixing the total strangeness $S$ and the total electric charge
$Q$, $\widetilde{\mu}_{S}$ and $\widetilde{\mu}_{Q}$ can be
expressed through baryonic potential $\widetilde{\mu}_{B}$ using
Eqs.~(\ref{D16}) and (\ref{D17}). Therefore the mean numbers of each
particle and resonance species at chemical freeze-out are determined
solely by the temperature $T$ and baryonic chemical potential
$\widetilde{\mu}_B$.

In practical calculations, we use the phenomenological observation
\cite{Cleymans05} that particle yields in central Au+Au or Pb+Pb
collisions in a wide center-of-mass energy range $\sqrt s_{NN}
=2.2-200$~GeV can be described within the thermal statistical
approach using the following parametrizations of the temperature and
baryon chemical potential \cite{Cleymans05}:
\begin{equation}
\label{Cleymans2}
T(\widetilde{\mu}_{B}) = a - b\widetilde{\mu}^2_B - c\widetilde{\mu}^4_B,
\end{equation}

\begin{equation}
\label{Cleymans1}
\widetilde{\mu}_{B}(\sqrt s_{NN}) = \frac{d}{1+e\sqrt s_{NN}},
\end{equation}
$a = 0.166 \pm 0.002$~GeV, $b = 0.139 \pm 0.016$~GeV$^{-1}$,
$c = 0.053 \pm 0.021$~GeV$^{-3}$ and
$d = 1.308 \pm 0.028$~GeV, $e = 0.273 \pm 0.008$~GeV$^{-1}$.

The particle densities at the chemical freeze-out
stage are too high (see, e.g., \cite{Sinyukov02})
to consider particles as free streaming and associate this stage
with the thermal freeze-out one.
The mean particle numbers $\bar{N}_i^{\rm th}$ at thermal
freeze-out can be determined using the following procedure \cite{Sinyukov02}.
First, the temperature and chemical
potentials at chemical freeze-out have to be fitted
from the ratios of the numbers of (quasi)stable particles.
The fitting procedure should account for the decays of all
resonances as well as unstable particles in given experimental conditions
(feed-down). The common factor, $V_{\rm eff}^{\rm ch}$,
and, thus, the absolute particle and resonance numbers can be fixed, e.g.,
from pion multiplicities. Within the concept of chemically
frozen evolution these numbers are
assumed to be conserved
except for corrections due to decay of some
part of short-lived resonances that can be estimated from the assumed
chemical to thermal freeze-out evolution time.
Then one can calculate the mean numbers of different
particles and resonances reaching a (common) thermal freeze-out
hypersurface. At a given thermal freeze-out temperature $T_{\rm th}$ these mean
numbers can be expressed through the thermal effective volume $V_{\rm eff}^{\rm th}$
and the chemical
potentials for each particle species $\mu^{{\rm th}}_{i}$ which
can no more be
expressed in the form $\mu_i =\vec{q}_i\vec{\widetilde{\mu}}$ valid only for
chemically equilibrated systems. For a given parametrization of the thermal
freeze-out hypersurface, the thermal effective volume $V_{\rm eff}^{\rm th}$
(and thus all $\mu^{{\rm th}}_{i}$)
can be fixed with the help of pion interferometry data.

In practical calculations we determine all macroscopic
characteristics of a particle system with the temperature $T$ and
chemical potentials $\mu_i$ via a set of equilibrium distribution
functions in the fluid element rest frame:
\begin{equation}
\label{GC1} f_{i}^{\rm eq}(p^{*0};T,\mu_{i})
 = \frac{1}{(2\pi)^3}\frac{g_i}{\exp{([p^{*0} - \mu_i]/T)}
\pm 1}.
\end{equation}
Eq.~$(\ref{M6})$  for the  particle number density then reduces to
\begin{equation}
\label{GC2} \rho_{i}^{\rm eq}(T, \mu_i) =  4\pi \int_{0}^{\infty} {dp^{*}
p^{*2} f_{i}^{\rm eq}(p^{*0}; T,\mu_i)}.
\end{equation}
Using the expansion
\begin{equation}
\label{GC3} f_i^{\rm eq}(p^{*0}; T,\mu_i) =
\frac{g_i}{(2\pi)^3}\sum_{k=1}^{\infty}(\mp)^{k+1}
\exp(k\frac{\mu_i - p^{*0}_i}{T}),
\end{equation}
the density can be represented in a form of a fast converging series:
\begin{equation}
\label{GC4} \rho_{i}^{\rm eq}(T, \mu_i) = \frac{g_i}{2 \pi^2}m^2_iT\sum_{k=1}^{\infty}
\frac{(\mp)^{k+1}}{k} \exp(\frac{k\mu_i}{T})K_2(\frac{km_i}{T}),
\end{equation}
where $K_2$ is the modified Bessel function of the
second order.

We assume that the calculated mean particle numbers
$\bar{N}_i=\rho_{i}^{\rm eq}V_{\rm eff}$ correspond to a grand
canonical ensemble. The probability that the ensemble consists of
$N_i$ particles is thus given by Poisson distribution:
\begin{equation}
\label{GC5} P(N_i) = \exp{(-\bar{N}_i)}\frac{(\bar{N}_i)^{N_i}}{N_i!}.
\end{equation}

\maketitle
\section{\label{sec3} Hadron momentum distributions}

We suppose that a hydrodynamic expansion of the fireball ends by
a sudden system breakup at given temperature and chemical
potentials. In this case
the momentum distribution of the produced hadrons
keeps the thermal character of the equilibrium distribution
$(\ref{M2P})$.
Similar to Eqs.~(\ref{M1}), (\ref{M2}), this
distribution
is then calculated according to the Cooper-Frye
formula \cite{Cooper74}:
\begin{equation}
\label{M71} p^0\frac{d^3\bar{N}_{i}}{d^3p} = \int_{\sigma(x)}
d^3\sigma_{\mu}(x) p^{\mu} f^{\rm eq}_i(p \cdot u(x);T,\mu_i).
\end{equation}
The integral in Eq.~($\ref{M71}$) can be calculated with the help of the
invariant weight
\begin{equation}
\label{M72} W_{\sigma, i}(x,p) \equiv p^0
\frac{d^6\bar{N}_i}{d^3\sigma d^3\vec{p}}= n_{\mu}(x) p^{\mu}
f_{i}^{\rm eq}(p \cdot u(x);T,\mu_i).
\end{equation}
It is convenient to transform the four-vectors into the fluid element rest
frame, e.g.,
\begin{equation}
\begin{array}{c}
\label{M72P}
n^{*0} = n^{\mu}u_{\mu}= \gamma(n^0 - \vec{v}\vec{n}),
\\
\vec{n}^{*} = \vec{n} - \gamma (1 + \gamma)^{-1}(n^{*0} +
n^0 )\vec{v}
\end{array}
\end{equation}
and calculate the weight in this frame:
\begin{equation}
\label{HD4}
W_{\sigma, i}(x,p) = W_{\sigma, i}^{*}(x^*,p^*) =
n_{\mu}^{*}(x) p^{*\mu} f_{i}^{\rm eq}(p^{*0};T,\mu_i).
\end{equation}

Particulary, in the case when the normal four-vector $n^{\mu}(x)$
coincides with the fluid element flow velocity $u^{\mu}(x)$, i.e.
$n^{*\mu}=u^{*\mu}=\{1,0,0,0\}$, the weight $W_{\sigma, i}^{*}(x^*,p^*) =
p^{*0} f_{i}^{\rm eq}(p^{*0};T,\mu)$ is independent of $x$ and isotropic in the
three-momentum ${\vec{p}}^{~*}$. A simple and $100 \%$ efficient simulation
procedure can then be realized in this frame and the four-momenta of the
generated particles transformed back to the fireball rest frame using the
velocity field $\vec{v}(x)$: \begin{equation}
\begin{array}{c}
\label{M69} p^{0} = \gamma(p^{0*} + \vec{v}\vec{p}^{~*}),
\\
\vec{p} = \vec{p}^{~*} + \gamma (1 + \gamma)^{-1} (p^{0*} +
p^0 )\vec{v}.
\end{array}
\end{equation}
There are two well-known examples of the models giving $n^{\mu}(x)=u^{\mu}(x)$:
the Bjorken model with hypersurface $\tau_B=(t^2-z^2)^{1/2}=const$ and absent transverse flow and the
model with hypersurface $\tau_H=(t^2-x^2-y^2-z^2)^{1/2}=const$ and spherically symmetric Hubble flow.
In general case $n_{\mu}(x)$ may differ from $u_{\mu}(x)$ and one should
account for the $x-p$ correlation and the corresponding anisotropy
due to the factor $n_{\mu}p^{\mu}$
even in
the fluid element rest frame \cite{GorSin}.

\maketitle
\section{\label{sec4}Generalization of the Cooper-Frye prescription}
It is well known that the Cooper-Frye freeze-out prescription in
Eq.~($\ref{M71}$) is not valid for the part of the freeze-out hypersurface
characterized by a space-like normal four-vector $n^{\mu}$.
In this case $|n^0|<|\vec{n}|$ and so $p^{\mu}n_{\mu}<0$
for some particle momenta thus leading to negative contributions to particle
numbers. Usually, the negative contributions are simply rejected
\cite{Sin4,Bugaev}.
This procedure however violates the continuity
condition of the flow $\rho_iu^{\mu}n_{\mu}$ through
the freeze-out hypersurface. Taking into account the continuity of the
particle flow, the generalization of
Eq.~($\ref{M71}$) has the form \cite{Sin4}:
\begin{equation}
\label{GCF} p^0\frac{d^3\bar{N}_{i}}{d^3p} = \int_{\sigma(x)}
d^3\sigma_{\mu}(x)\pi^{\mu}(x,p) f_{i}^{\rm eq}(T(x),\mu_i(x)),
\end{equation}
where
\begin{equation}
\begin{array}{c}
\label{GCFPI}
\pi^{\mu}(x,p)=p^{\mu}\theta(1-|\widetilde{\lambda}(x,p)|)+
u^{\mu}(x)~p \cdot u(x)~ \theta(|\widetilde{\lambda}(x,p)|-1),
\\
\widetilde{\lambda}(x,p)=1-{p \cdot n(x)}~[{p \cdot u(x)~ n(x) \cdot
u(x)}]^{-1},
\end{array}
\end{equation}
$\theta(x)=1$ for $x\ge 0$, $\theta(x)=0$ for $x < 0$.

Passing to the fluid element rest frames at each point $x$
and using Lorentz transformation properties of the quantities in
Eq.~($\ref{GCF}$),
one arrives at the same form of the four-vector of particle flow
as in the case of the freeze-out hypersurface with the time-like normal
$n^{\mu}(x)$:
\begin{equation}
\label{GCFJ} j^{\mu}(x)=\int \frac{d^3\vec{p}}{p_0}\pi^\mu(x,p)f_{i}^{\rm
eq}(T(x),\mu_i(x))=\rho_i^{\rm eq}(T(x), \mu_i(x))u^{\mu}(x).
\end{equation}

Therefore the factorization of the freeze-out details in the effective volume
in the case of constant temperature and chemical potentials, i.e.
Eq.~($\ref{M8}$), is valid for any type of hypersurface \cite{AkkSin1}.
 It follows from Eqs.~($\ref{GCF}$), ($\ref{GCFPI}$)
that the invariant weight in the fluid element rest frame has then the form:
\begin{equation}
\label{GCFW}
W^{*}_{\sigma,i}(x^*,p^{*}) =
\left[p^{* \mu}n_{\mu}^{*}
~\theta\left(1-\left|\frac{\vec{p}^{~*}\vec{n}^{~*}}{p^{*0}n^{*0}}\right|\right)
+p^{*0}n^{*0}~\theta\left(\left|
\frac{\vec{p}^{~*}\vec{n}^{~*}}{p^{*0}n^{*0}}\right|-1\right)\right]
f_{i}^{\rm eq}(p^{*0};T,\mu_i).
\end{equation}
For the time-like normal $n^{\mu}(x)$,
Eq.~(\ref{GCFW}) reduces to Eq.~(\ref{HD4}).

It is worth noting that though the bulk of particles is likely
associated with the volume decay, the
particle emission from the surface of expanding system, or formally,
from a non-space-like part of the freeze-out hypersurface enclosed
in Minkowski space, is essential for a description of hadronic
spectra and like pion correlations at relatively large $p_T$
\cite{Borysova}.

\maketitle
\section{\label{sec6}Freeze-out surface parameterizations}
In principle, one can specify the fireball initial conditions (e.g.,
Landau- or Bjorken-like) and equation of state to follow the fireball
dynamic evolution until the freeze-out stage with the help of relativistic
hydrodynamics. The corresponding freeze-out four-coordinates $x^\mu$, the
hypersurface normal four-vectors $n^\mu(x)$ and the collective flow
four-velocities $u^\mu(x)$ can then be used to calculate particle spectra
according to generalized Cooper-Frye prescription. This possibility is
forseen as an option in our MC generator. In this paper, we however do not
consider the fireball evolution, we demonstrate our fast MC procedure
utilizing the simple and frequently used parametrizations of the
freeze-out.

At relativistic energies, due to dominant longitudinal motion,
it is convenient to substitute the Cartesian coordinates $t$, $z$ by
the Bjorken ones
\begin{equation}
\label{FS2}
\tau = (t^2 - z^2)^{1/2},~~
\eta = \frac{1}{2} \ln \frac{t+z}{t-z}
\end{equation}
and introduce the the radial
vector $\vec{r}\equiv\{x,y\}=\{r \cos \phi, r \sin \phi\}$, i.e.:
\begin{equation}
\label{xB}
x^{\mu} =
\{\tau \cosh \eta , \vec{r} , \tau \sinh\eta\}=
\{\tau \cosh \eta ,r \cos \phi, r \sin \phi , \tau \sinh\eta\}.
\end{equation}
Similarly, it is convenient to parameterize the fluid
flow four-velocity $u^{\mu}(x)= \gamma(x) \{1,\vec{v}(x)\}
\equiv \gamma(x) \{1,\vec{v}_r(x),v_z(x)\}$
at a point $x$ in terms of the
longitudinal ($z$) and transverse ($r$) fluid flow rapidities
\begin{equation}
\label{HC2}\eta_u(x) = \frac{1}{2}
\ln{\frac{1+v_z(x)}{1-v_z(x)}},~~
\rho_u(x) = \frac{1}{2}
\ln{\frac{1+v_r(x)\cosh\eta_u(x)}{1-v_r(x)\cosh\eta_u(x)}},
\end{equation}
where
$v_{r} = | \vec{v}_{r}|$ is the magnitude of the transverse component of the flow
three-velocity
$ \vec{v}=
\{ v_r \cos \phi_u , v_r \sin \phi_u, v_z\}$,
i.e.
\begin{equation}
\begin{array}{c}
\label{HC6}u^{\mu}(x)=
\{\cosh \rho_u \cosh \eta_u, \sinh \rho_u \cos \phi_u, \sinh \rho_u \sin \phi_u,
\cosh \rho_u \sinh \eta_u \}
\\
=\{
 (1+u_{r}^{2})^{1/2} \cosh \eta_u,\vec{u}_r, (1+u_{r}^{2})^{1/2} \sinh \eta_u\},
\end{array}
\end{equation}
$\vec{u}_{r} = \gamma\vec{v}_r= \gamma_{r}\cosh\eta_u \vec{v}_r$,
$\gamma_r=\cosh \rho_u$.
For the considered central collisions of symmetric nuclei, $\phi_u=\phi$.
Representing the
freeze-out hypersurface by the equation $\tau=\tau(\eta,r,\phi),$ the
hypersurface element in terms of the coordinates $\eta$, $r$,
$\phi$ becomes
\begin{equation}
\label{FS1}
d^3\sigma_{\mu}= \epsilon_{\mu
\alpha \beta \gamma} \frac{dx^{\alpha}dx^{\beta}dx^{\gamma}} {d\eta dr
d\phi}d\eta dr d\phi,
\end{equation}
where $\epsilon_{\mu \alpha \beta\gamma}$ is the completely
antisymmetric Levy-Civita tensor in four dimensions with
$\epsilon^{0123} = -\epsilon_{0123} = 1$. Particulary, for
azimuthaly symmetric hypersurface $\tau=\tau(\eta,r)$,
Eq.~($\ref{FS1}$) yields \cite{Sinyukov02}:
\begin{equation}
\label{FS4}
d^3\sigma_{\mu}=\tau(\vec{r},\eta)d^2\vec{r}d\eta
\{\frac{1}{\tau}\frac{d\tau}{d\eta}\sinh\eta
+ \cosh\eta,-\frac{d\tau}{dr}\cos\phi,-\frac{d\tau}{dr}\sin\phi,
-\frac{1}{\tau}\frac{d\tau}{d\eta}\cosh\eta -\sinh\eta\}.
\end{equation}
Generally, the freeze-out hypersurface is represented by a set of equations
$\tau=\tau_{j}(\eta, r, \phi)$ and Eq.~($\ref{FS1}$) should be substituted by
the sum of the corresponding hypersurface elements.

To simplify the situation, besides the azimuthal symmetry, we further assume
the longitudinal boost invariance \cite{Bjorken83}. The local quantities (such
as particle density) are then functions of $\tau$ and ${r}$ only. The
hypersurface then takes the form $\tau=\tau(r)$,  the flow rapidities
$\eta_u = \eta$ (i.e. $v_z=z/t$), $\rho_u=\rho_u(r)$ and Eq.~($\ref{FS4}$)
yields
\begin{equation}
\label{FS5}
\begin{array}{c}
d^3\sigma_{\mu}=\tau(r)d^2\vec{r}d\eta
\{\cosh\eta,-\frac{d\tau}{dr}\cos\phi,-\frac{d\tau}{dr}\sin\phi,
-\sinh\eta \},
\cr
d^3\sigma=|1-(\frac{d\tau}{dr})^{2}|^{1/2}\tau(r)d^2\vec{r}d\eta,
\cr
n^{\mu}(x) =|1-(\frac{d\tau}{dr})^{2}|^{-1/2}\{
\cosh\eta,\frac{d\tau}{dr}\cos\phi,\frac{d\tau}{dr}\sin\phi,
\sinh\eta \}.
\end{array}
\end{equation}
Note that the normal four-vector $n^{\mu}$ becomes space-like
$(n^{\mu}n_{\mu}=-1)$ for $|d\tau/dr|>1$.

For the simplest freeze-out hypersurface $\tau = const$
one has
\begin{equation}
\begin{array}{c}
\label{d3sigmab}
    d^3\sigma=\tau d^2\vec{r}d\eta,
\\
    n^{\mu}(x)=\{\cosh \eta, 0, 0, \sinh \eta \}.
\end{array}
\end{equation}
In this case the normal $n^{\mu}(x)$ is time-like ($n^{\mu}n_{\mu}=1$)
but generally different from the flow four-velocity $u^{\mu}(x)$
except for the case of absent transverse flow (i.e. $\rho_u=0$).
Assuming $\phi_u=\phi$ and
the linear transverse flow rapidity
profile (effectively taking into account a positive flow - radius
correlation up to the radii close to the fireball boundary as
indicated by numerical solutions of
(3+1)-dimensional relativistic hydrodynamics, see, e.g., \cite{mor02}):
\begin{equation}
\label{URBa}
    \rho_u=\frac{r}{R}\rho_u^{\max},
\end{equation}
where $R$ is the fireball transverse radius,
the total effective volume for particle
production at $\tau = const$ is
\begin{eqnarray}
\label{HD3} V_{\rm eff}=\int_{\sigma(x)} d^3\sigma_{\mu}(x)u^{\mu}(x) =
\tau \int_0^{R} \gamma_r r dr \int_0^{2\pi} d\phi
\int_{\eta_{\min}}^{\eta_{\max}} d\eta =\nonumber \\
=2\pi\tau\Delta\eta\left({R\over\rho_u^{\max}}\right)^2
(\rho_u^{\max}\sinh \rho_u^{\max}-\cosh \rho_u^{\max}+1),
\end{eqnarray}
where $\Delta\eta = \eta_{\max} - \eta_{\min}$.
For small values of the maximal transverse flow rapidity $\rho_u^{\max}$,
Eq.~(\ref{HD3}) reduces to
$V_{\rm eff}=\pi\tau R^2\Delta\eta$ \cite{Sinyukov02}.

We shall refer the above choice of the freeze-out hyper-surface
and the flow four-velocity profile as the Bjorken-like parametrization or
Bjorken model scenario for particle freeze-out with transverse flows
\cite{Bjorken83}.

We also consider so called Cracow model scenario \cite{Bron04}
corresponding to the Hubble-like freeze-out hypersurface
$\tau_H=(t^2-x^2-y^2-z^2)^{1/2}=const$
and flow four-velocity
\begin{equation}
\label{URH}
u^{\mu}(x)=x^{\mu}/\tau_H.
\end{equation}
Introducing the longitudinal space-time rapidity $\eta$ according to Eq.~(\ref{FS2})
and the transverse space-time rapidity $\rho=\sinh^{-1}(r/\tau_H)$,
one has \cite{Csorgo96}
\begin{equation}
\label{xH}
x^{\mu}=\tau_H\{\cosh \eta \cosh \rho, \sinh \rho \cos \phi, \sinh \rho
\sin \phi, \sinh \eta \cosh \rho \},
\end{equation}
$\tau_H=\tau_B/\cosh \rho$.
Representing the freeze-out hypersurface by the equation
$\tau_H=\tau_H(\eta,\rho,\phi)=const$, one finds from
Eq.~($\ref{FS1}$):
\begin{equation}
\begin{array}{{c}}
\label{cracow}
d^3\sigma=\tau^{3}_H\sinh \rho \cosh \rho d \eta d \rho d \phi =\tau_H
d \eta d^2\vec{r},
\\
n^{\mu}(x)=u^{\mu}(x).
\end{array}
\end{equation}
The effective volume corresponding to $r=\tau_H \sinh \rho<R$
and $\eta_{\min} \le \eta \le \eta_{\max}$
is
\begin{equation}
\label{HD5}
V_{\rm eff}=\int_{\sigma(x)}d^3 \sigma_{\mu}(x)u^{\mu}(x) = \tau_H
\int^{R}_{0} r dr \int_{0}^{2 \pi} d \phi \int_{\eta_{\min}}^{\eta_{\max}}d \eta
= \pi \tau_H R^{2} \Delta\eta.
\end{equation}

\maketitle
\section{\label{sec7}Hadron generation procedure}

Our MC procedure to generate the freeze-out
hadron multiplicities, four-momenta and four-coordinates is the following:

\begin{enumerate}

\item
First, the parameters of the chosen freeze-out model are
initialized.
Particularly, for the models with constant
freeze-out temperature  $T$ and chemical potentials $\mu_i$,
the phenomenological
formulae $(\ref{Cleymans2})$, $(\ref{Cleymans1})$ are implemented
as an option allowing to calculate
$T$ and $\mu_i$ at the chemical freeze-out
in central $Au+Au$ or $Pb+Pb$ collisions
specifying only the center-of-mass energy $\sqrt s_{NN}$.
In the scenario with the thermal freeze-out occurring
at a temperature $T^{\rm th} < T^{\rm ch}$,
the chemical potentials $\mu_i^{\rm th}$
are no more given by Eq.~(\ref{mui}).
At given thermal freeze-out
temperature $T^{\rm th}$ and effective volume
$V_{\rm eff}^{\rm th}$,
they are set according to the
procedure described in section \ref{sec2}.
So far, only the stable particles and resonances
consisting of $u$, $d$, $s$ quarks
are incorporated in the model. They are taken from the
ROOT particle data table \cite{ROOT,pdg}.

\item
Next, the effective volume corresponding to a given freeze-out
model is determined, e.g., according to Eq.~(\ref{HD3}) or
(\ref{HD5}) and particle number densities are calculated with the
help of Eq.~(\ref{GC4}). The mean multiplicity of each particle
species is then calculated according to Eq.~(\ref{M8}).
A more general option to calculate the mean multiplicities, e.g., in the case
of the freeze-out hypersurface obtained from relativistic hydrodynamics, is
the direct integration of Eq.~($\ref{GCF}$).
The multiplicity corresponding to the mean one is simulated according to Poisson
distribution in Eq.~(\ref{GC5}).

\item  The particle freeze-out four-coordinates 
$x^\mu=\{\tau\cosh\eta,r\cos\phi,r\sin\phi,\tau\sinh\eta\}$
in the fireball rest frame are then simulated on each hypersurface segment
$\tau=\tau_j(r)$ according to the element $d^3\sigma_\mu u^\mu =
d^3\sigma_0^* = n_0^*(r) |1-(d\tau/dr)^2|^{1/2}\tau(r)d^2\vec{r}d\eta$,
assuming $n_0^*$ and $\tau$ functions of
$r$ (i.e. independent of $\eta, \phi$),
by sampling uniformly distributed $\eta$ in the interval
$[\eta_{\min},\eta_{\max}]$, $\phi$ in the interval $[0, 2\pi]$ and generating
$r$ in the interval $[0, R]$) using a $100 \%$ efficient
procedure similar to ROOT routine $GetRandom()$.
In the Bjorken- and Hubble-like models: $\tau(r)=\tau_B=const$,
$n_0^*=\cosh\rho_u=\gamma_r$ and
$|1-(d\tau/dr)^2|^{1/2}\tau(r)=\tau_H=const$, $n_0^*=1$, respectively.
Note that if $n_0^*$ and $\tau$ were
depending on two or three variables,
a generalization of the routine $GetRandom()$ to more dimensions
is possible. 
A less efficient possibility is to simulate $\vec{r}, \eta$ according to
the element $d^2\vec{r}d\eta$ and include the factor
$d^3\sigma n_0^*/d^2\vec{r}d\eta$ in the residual weight in the step 6.
Also note that the particle freeze-out coordinates calculated from
relativistic hydrodynamics are distributed according to the element
$d^3\sigma_{\mu} u^\mu$.

\item The corresponding collective flow four-velocities $u^{\mu}(x)$
are calculated using, e.g., Eqs. (\ref{HC6}), (\ref{URBa})
or Eq.~(\ref{URH}).

\item The particle three-momenta
$p^{*}\{\sin\theta\cos\phi,\sin\theta\sin\phi,\cos\theta\}$
are then generated in the fluid element rest
frames according to the probability
 $f_i^{\rm eq}(p^{0*};T,\mu_i)p^{*2}dp^{*}d\cos\theta_p^* d\phi_p^*$
by sampling uniformly distributed $\cos\theta_p^*$ in the interval
$[-1,1]$, $\phi_p^*$ in the interval $[0, 2\pi]$ and generating
$p^*$ using a $100 \%$ efficient
procedure similar to ROOT routine $GetRandom()$.

\item
Next, the standard von Neumann rejection/acceptance procedure is
used to account for the difference between the true probability
$W_{\sigma,i}^* d^3\sigma d^3\vec{p}^{~*}/p^{0*}$
(see Eqs. (\ref{M72}), (\ref{HD4}), (\ref{GCFW}))
and the probability $f_i^{\rm eq}(p^{0*};T,\mu_i)d^3\sigma_{\mu} u^\mu d^3\vec{p}^*
=f_i^{\rm eq}(p^{0*};T,\mu_i)n^{0*}d^3\sigma d^3\vec{p}^*$
corresponding to the simulation steps 3-5.
Thus the residual weight
\begin{equation}
\label{wres1}
W_i^{\rm res}=\frac{W_{\sigma,i}^*
d^3\sigma d^3\vec{p}^* }
{n^{0*}p^{0*}f_i^{\rm eq} d^3\sigma d^3\vec{p}^*}
\end{equation}
is calculated and the simulated particle four-coordinate and
four-momentum are accepted provided that this weight is larger than
a test variable randomly simulated in the interval
$[0,\max(W_i^{\rm res})]$. Otherwise, the simulation returns to
step 3. Note that for the freeze-out parametrizations considered
in this paper,
\begin{equation}
\label{wres2}
W_i^{\rm res}=\left(1-\frac{\vec{n}^{~*}\vec{p}^{~*}}{n^{0*}p^{0*}}\right)
\end{equation}
and the maximal weight $\max(W_i^{\rm res})$ can be calculated analytically.
Particularly, in the Bjorken-like model and $\eta^{\max}\gg 1$,
$W_i^{\rm res}$ is distributed in the interval
$[1-\tanh\rho_u^{\max},1+\tanh\rho_u^{\max}]$.
The step 6 can be omitted for the Hubble-like model or
for the Bjorken model without transverse flow
($\rho_u=0$) when
$W_i^{\rm res}=1$.
Generally, in the residual weight one should take
into account the contribution of non-space-like sectors of
the freeze-out hypersurface:
\begin{equation}
\label{wres3}
W_i^{\rm res}=\left[\left(1-\frac{\vec{n}^{~*}\vec{p}^{~*}}{n^{0*}p^{0*}}\right)
~\theta\left(1-\left|\frac{\vec{p}^{~*}\vec{n}^{~*}}{p^{*0}n^{*0}}\right|\right)
+ \theta\left(\left|
\frac{\vec{p}^{~*}\vec{n}^{~*}}{p^{*0}n^{*0}}\right|-1\right)
\right].
\end{equation}

\item Next, the hadron four-momentum $p^{*\mu}$ is
boosted to the fireball rest frame according to Eqs.~(\ref{M69}).

\item
The two-body, three-body and many-body decays are simulated with
the branching ratios calculated via ROOT
utilities \cite{ROOT}.
A more correct kinetic evolution,
taking into account not only resonance decays but also hadron elastic scattering,
may be included with the help
of the Boltzmann equation solver C++ code
which was developed earlier \cite{UKM05}.

\end{enumerate}

It should be stressed that a high generation speed
is achieved due to $100 \%$ generation efficiency
of the freeze-out four-coordinates and four-momenta
in steps 3-5 as well as due to a weak
non-uniformity of the residual weight $W_i^{\rm res}$
in the cases of practical interest.
For example, in the Bjorken-like model, the increase
of the maximal transverse flow rapidity from zero ($W_i^{\rm res}=const$)
to $\rho_u^{\max}=0.65$ leads only to a few percent decrease of the
generation speed. Compared, e.g., to THERMINATOR \cite{THERMINATOR},
our generator appears more than one order of magnitude faster.

\maketitle
\section{\label{sec8} Validation of the MC procedure}

In the Boltzmann approximation for the equilibrium distribution function (\ref{GC1}),
i.e. retaining only the first term in the expansion (\ref{GC3}),
the transverse momentum ($p_t$) spectrum in the Bjorken-like model takes
the form \cite{Snigiriev96, heinz93}:
\begin{eqnarray}
\frac{d\bar{N}_{i}}{p_t dp_t} & = & \frac{1}{\pi}g_i ~\tau~ m_t~ e^{\mu_i/T}~\Delta\eta~
 \int_0^{R}r dr
K_1 \left(\frac{m_t\cosh{\rho_u}}{T}\right)
I_0\left(\frac{p_t\sinh{\rho_u}}{T} \right),
\label{ptspec}
\end{eqnarray}
where $I_0(z)$ and $K_1(z)$ are the
modified Bessel functions and  $m_t=(m^2_i+p_t^2)^{1/2}$ is the
particle transverse mass.

\begin{figure}
\begin{center}
\begin{tabular}{cc}
\includegraphics[width=8.3cm]{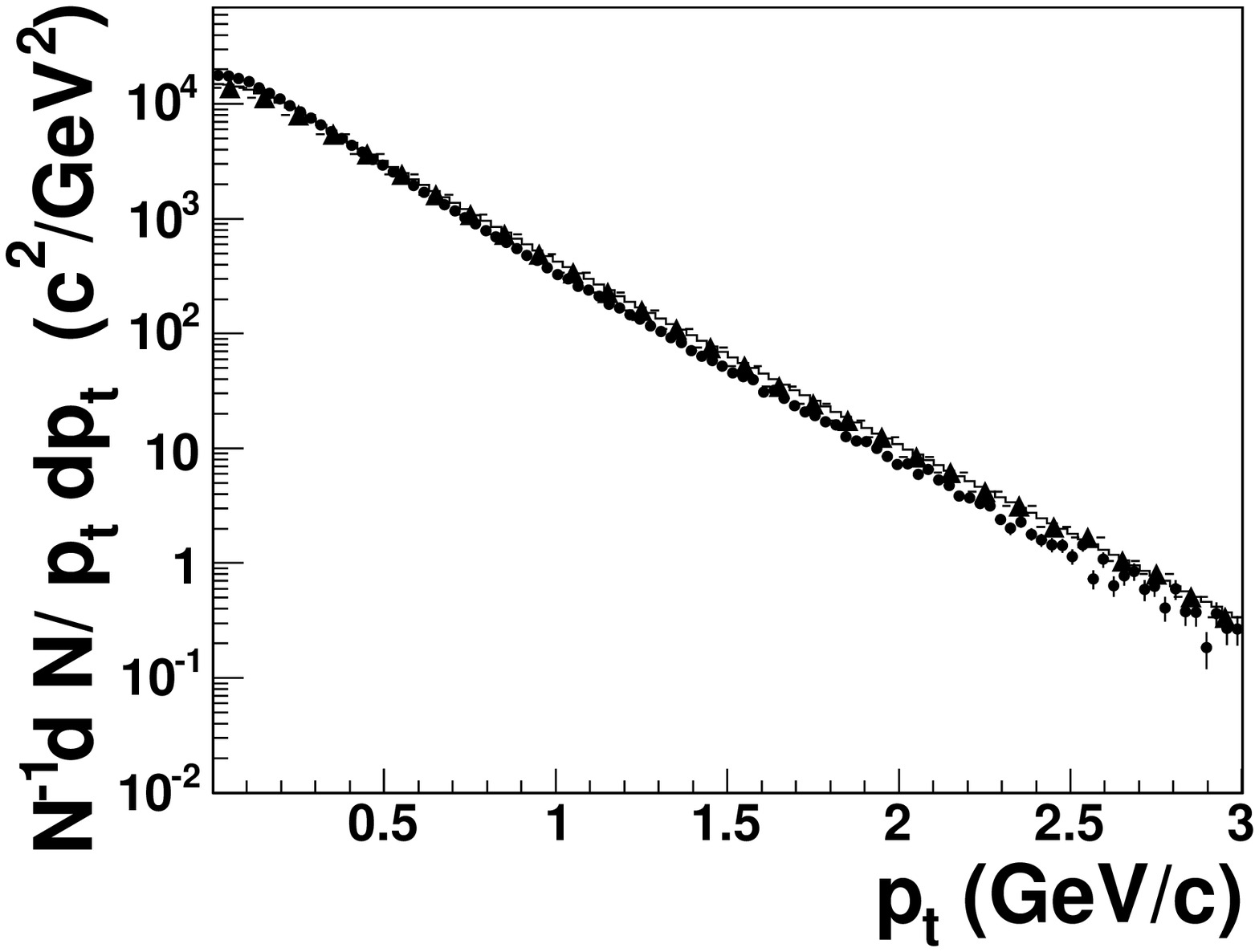}
\includegraphics[width=8.3cm]{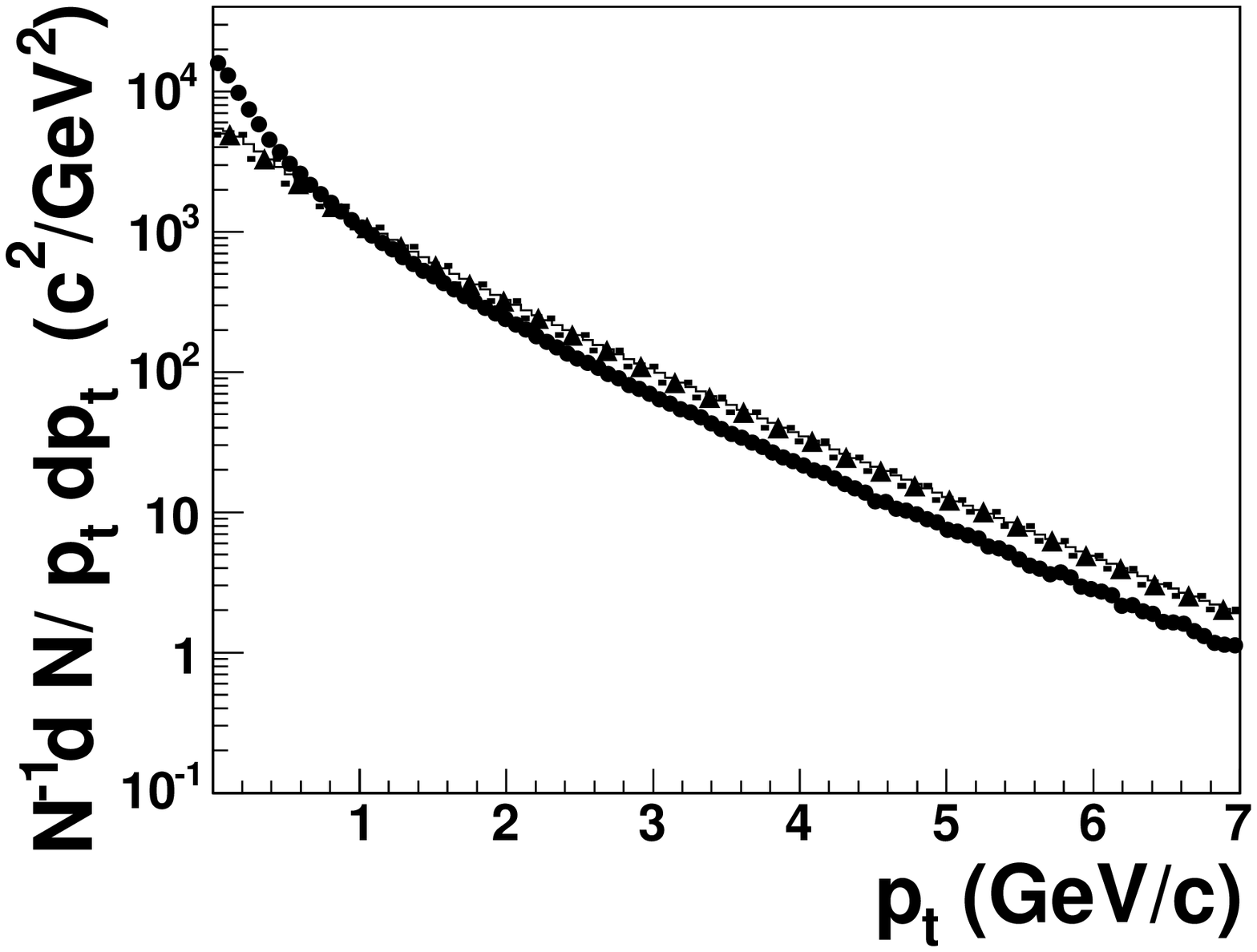}
\end{tabular}
\end{center}
\caption{The validation of the MC procedure for $\rho_u^{\max}=0.65$ (left panel)
and 2.0 (right panel):
the transverse momentum spectra (solid lines) calculated according to Eq.~(\ref{ptspec})
and the corresponding MC results (black triangles).
Also shown are the MC results obtained with a constant
residual weight (black points).
\label{fig:Exact_eq} }
\end{figure}

To test our MC procedure, we compare in
Fig.~\ref{fig:Exact_eq}
the transverse momentum spectrum calculated according to Eq.~(\ref{ptspec})
with the corresponding MC result for
$T=0.165$~GeV, $R=8$~fm,
$m_i=0.14$~GeV, $\Delta\eta=10$, $\mu=0.0$~GeV, $\tau=12$~fm/c,
$\rho_u^{\max}= 0.65$ and 2.0.
One may see that the analytical and the MC calculations practically coincide.

To demonstrate the increasing influence of the residual weight
with the increasing $\rho_u^{\max}$, we also present in Fig.~\ref{fig:Exact_eq}
the MC results obtained without this weight.

\maketitle
\section{\label{sec9}Input parameters and example calculations}
We present here the results of example MC calculations performed
on the assumption of a common chemical and
thermal freeze-out and compare them
with the experimental data on central
Au + Au collisions at RHIC.

\subsection{\label{sec86} Model input parameters}
First, we summarize the input parameters which control the
execution of our MC hadron generator
in the case of Bjorken-like and Hubble-like parametrizations
with a common thermal and chemical freeze-out:

\begin{enumerate}
\item Number of events to generate.
\item  Thermodynamic parameters at chemical freeze-out: temperature
($T$) and chemical potentials per a unit charge
$(\widetilde{\mu}_B, \widetilde{\mu}_S, \widetilde{\mu}_Q)$.
As an option, there is an additional parameter $\gamma_s \le 1$
taking into account the strangeness suppression according to partially
equilibrated distribution \cite{Gorenstein97,Raf81}:
\begin{equation}
f_{i}(p^{*0};T,\mu_i,\gamma_s) = \frac{g_i}{\gamma_s^{-n_i^s}\exp{([p^{*0} -
\mu_i]/T)} \pm 1},
\end{equation}
where $n_i^s$ is the number of strange quarks and
anti-quarks in a hadron $i$.
Optionally, the parameter $\gamma_s$ can be fixed using its
phenomenological dependence on the temperature and
baryon chemical potential \cite{Beca06}.
\item Volume parameters:  the freeze-out proper time ($\tau$)
and firebal transverse radius ($R$).
\item  Maximal transverse flow rapidity $(\rho_u^{\max})$ for Bjorken-like
parametrization \cite{Snigiriev96, Kruglov97}.
\item  Maximal space-time longitudinal rapidity ($\eta_{\max}$) which
determines the rapidity interval $[-\eta_{\max}, \eta_{\max}]$
in the collision center-of-mass system.
To account for the violation of the boost invariance, we have
included in the code an option corresponding to the substitution
of the uniform distribution of the space-time longitudinal
rapidity $\eta$ in the interval $[-\eta_{\max},\eta_{\max}]$
by a Gaussian distribution $\exp(-\eta^2/2\Delta\eta^2)$
with a width parameter $\Delta\eta$
(see, e.g., \cite{Wiedemann_Heinz02}).
\end{enumerate}

The parameters used to model central Au+Au collisions at
$\sqrt s_{NN} = 200$~GeV are given in Table \ref{tab:table4}.
\begin{table}
 \caption{\label{tab:table4}
The model parameters for central Au + Au
collisions at $\sqrt s_{NN} = 200$~GeV.
 }
 \begin{ruledtabular}
 \begin{tabular}{ccc}
 parameter       & Bjorken-like & Hubble-like\\
 \tableline
$T$, GeV            &   0.165   &   0.165 \\
$\widetilde{\mu}_{B}$, GeV   & 0.028 & 0.028  \\
$\widetilde{\mu}_{S}$, GeV    & 0.007 & 0.007  \\
$\widetilde{\mu}_{Q}$, GeV   & -0.001&-0.001  \\
$\gamma_s$        & 1 (0.8)  & 1 (0.8) \\
$\tau$, fm/c     & 6.1   & 9.65  \\
$R$,  fm         & 10.0 & 8.2 \\
$\eta_{\max}$        & 2 (3,5)  & 2 (3,5) \\
$\rho_u^{\max}$ & 0.65 & -  \\
\end{tabular}
 \end{ruledtabular}
 \end{table}

\subsection{\label{sec94}
Space-time distributions of the hadron
emission points}

\begin{figure}
\begin{center}
\begin{tabular}{cc}
\includegraphics[width=8.3cm]{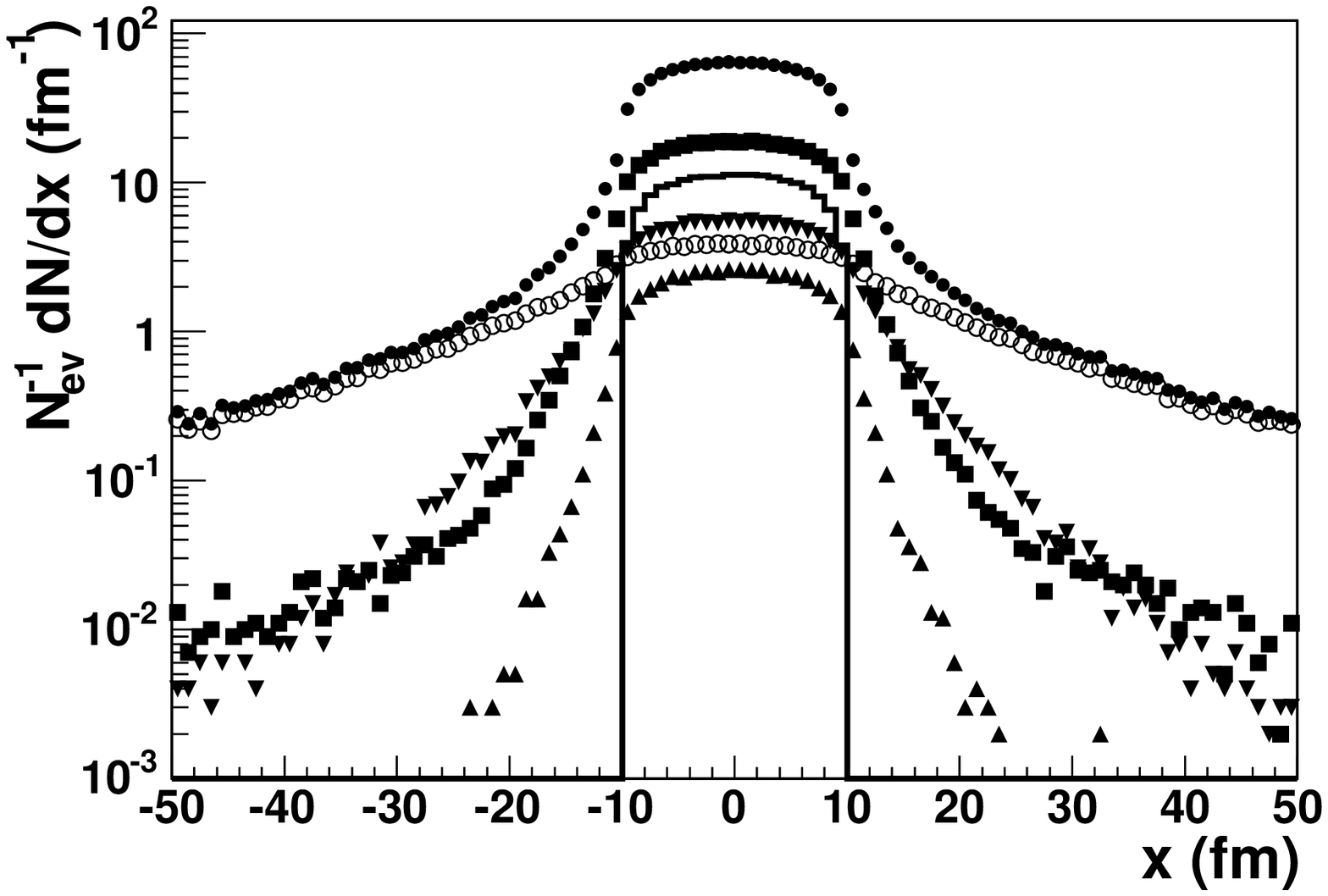}
\includegraphics[width=8.3cm]{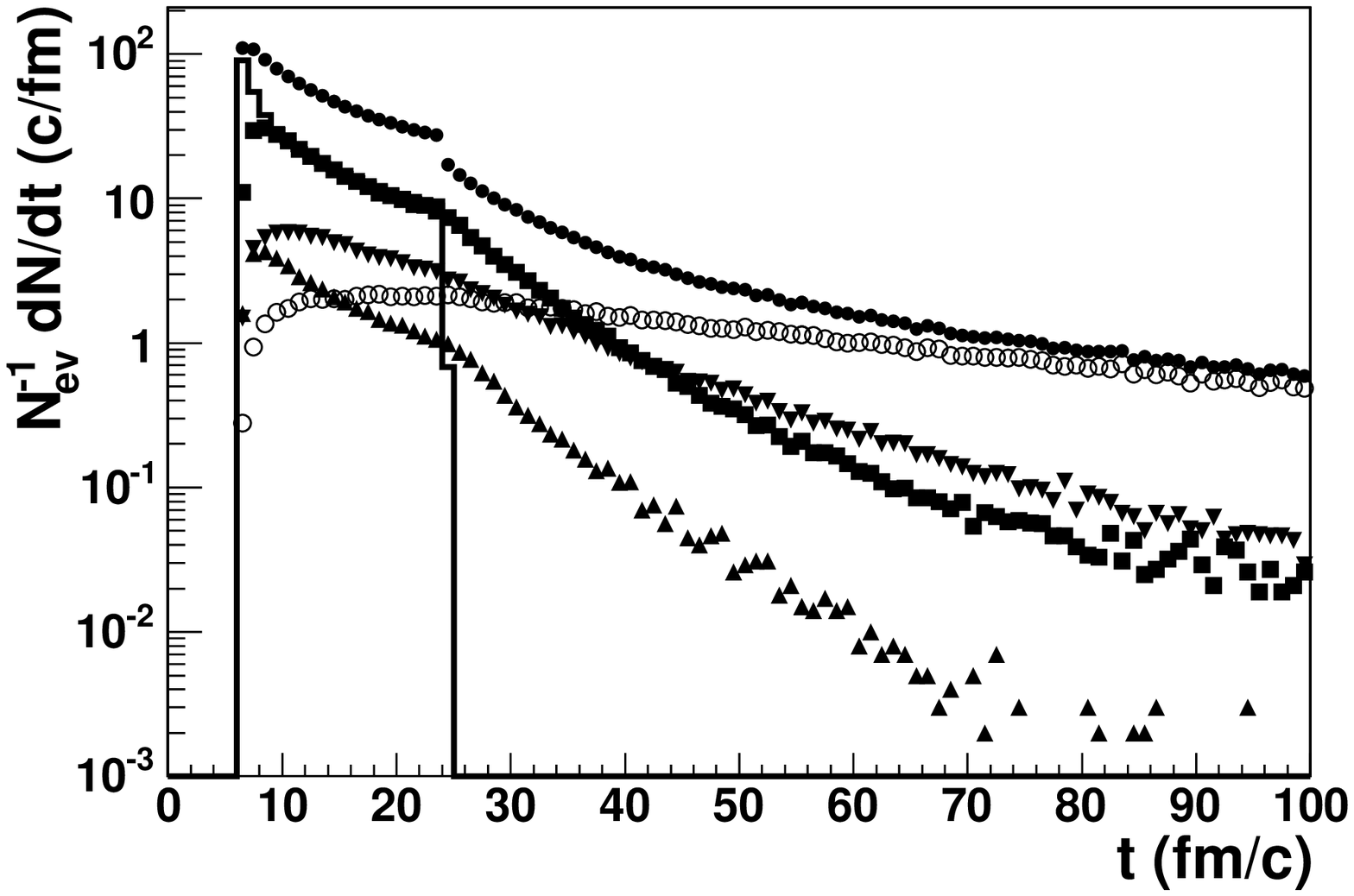}
\end{tabular}
\end{center}
\caption{The $\pi^+$ emission transverse x-coordinate (left)
and time (right) generated in the Bjorken-like model with the parameters
given in Table \ref{tab:table4}, $\eta_{\max}=2$:
all $\pi^+$'s (solid circles), direct $\pi^+$'s (solid line), decay $\pi^+$'s
from $\rho$ (squares), $\omega$ (open circles), $K^{*}(892)$ (up-triangles)
and $\Delta$ (down-triangles).
\label{fig:XTSL}
}
\end{figure}

\begin{figure}
\begin{center}
\begin{tabular}{cc}
\includegraphics[width=8.3cm]{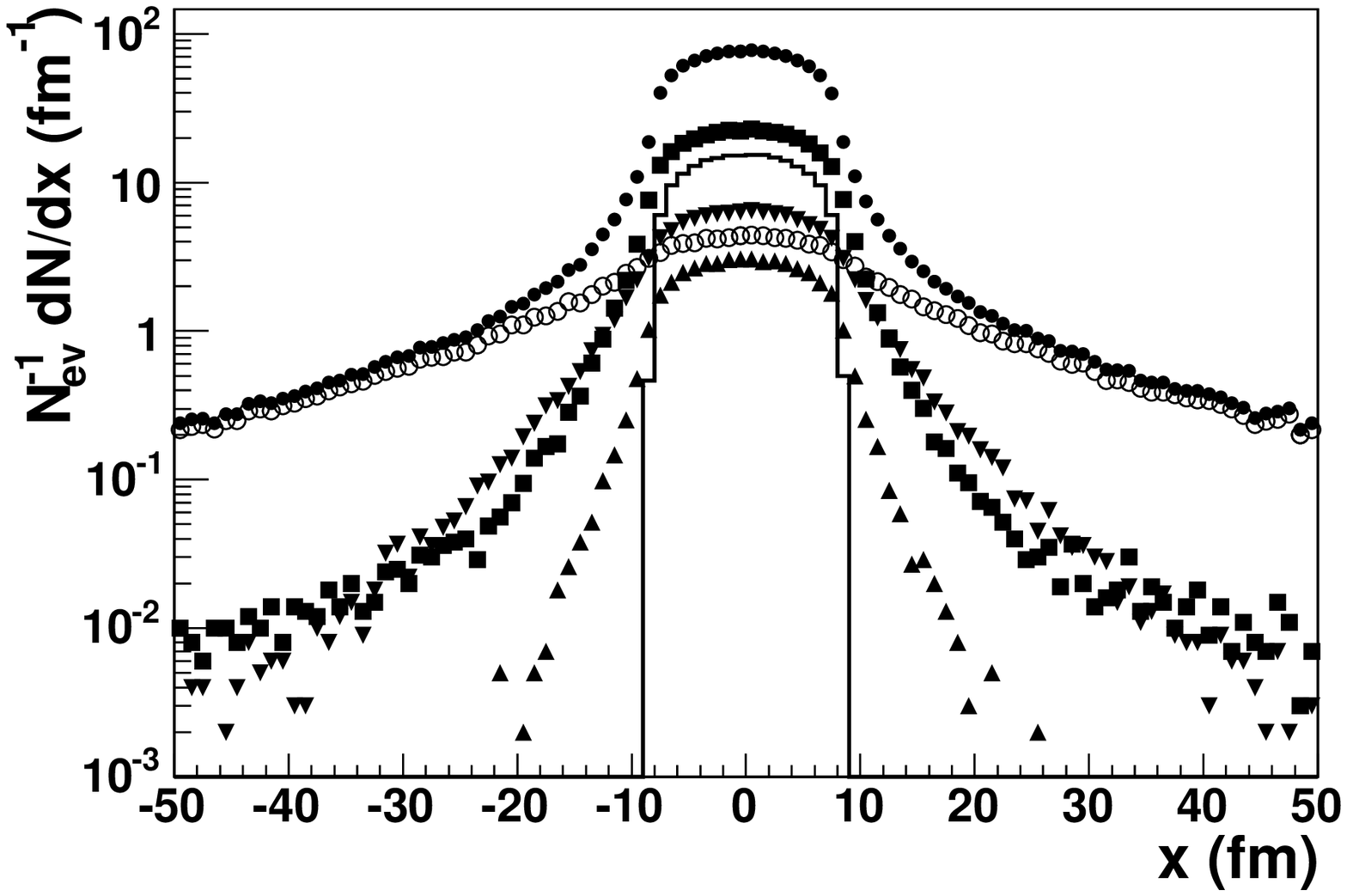}
\includegraphics[width=8.3cm]{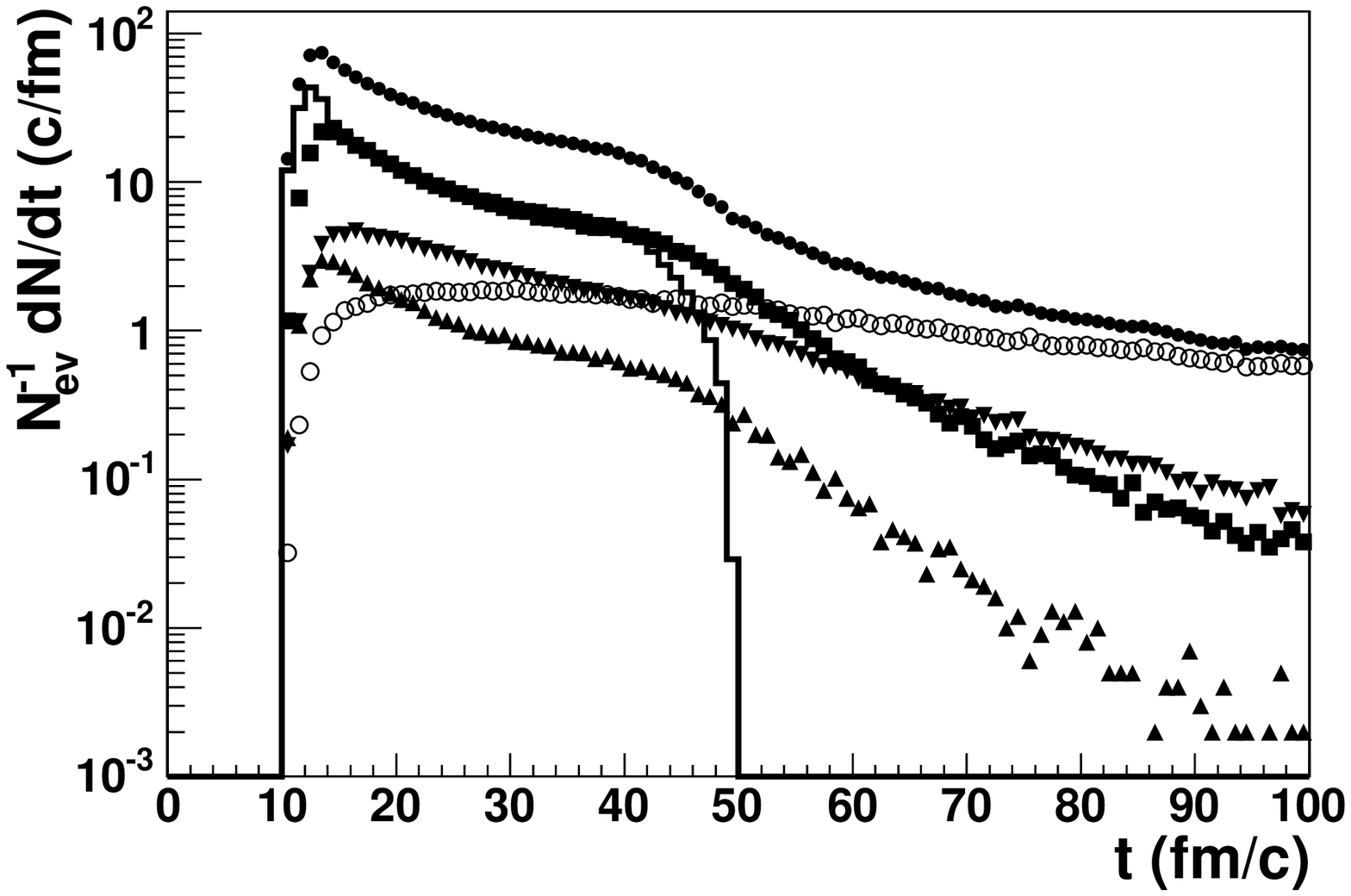}
\end{tabular}
\end{center}
\caption{The same as in Fig. \ref{fig:XTSL} for the Hubble-like parametrization.
\label{fig:XTH}
}
\end{figure}

In figures \ref{fig:XTSL} and \ref{fig:XTH},
we show the distributions of the $\pi^+$ emission transverse x-coordinate
and time generated in the Bjorken-like and Hubble-like models
with the parameters given in Table \ref{tab:table4}, $\eta_{\max}=2$.
Also shown are the contributions from the primary $\pi^+$'s emitted directly
from the freeze-out hypersurface and the contributions from $\pi^+$'s
from the decays of the most abundant resonances $\rho$, $\omega$,
$K^{*}(892)$ and $\Delta$.
For primary pions, $x<R$ and
$\tau<t<\tau \cosh\eta_{\max}$.
The tails at $|x|>R$ and $t>\tau \cosh\eta_{\max}$
reflect the exponential law of the resonance decays. The
longest tails in figures \ref{fig:XTSL} and \ref{fig:XTH}
are due to pions from $\omega$ decays.

\subsection{\label{sec91} Ratios of hadron abundances} \vskip 3mm
It is well known that the particle abundances
in heavy-ion collisions in a large energy range
can be reasonably well described within statistical models
(see, e.g., \cite{BHS95,Gorenstein97, THERMUS})
based on the assumption that the produced hadronic matter
reaches thermal and chemical equilibrium.
This is demonstrated in tables
\ref{tab:table1} and \ref{tab:table2} for
the particle number ratios near mid-rapidity in central Au +Au collisions
at $\sqrt s_{NN} = 130$ and 200~GeV calculated
in our MC model and the statistical model of Ref.~\cite{Bron02}
and compared with the RHIC experimental data.
Being independent of volume and flow parameters, the particle
number ratio allow one to fix the thermodynamic parameters.
We have not tuned the latter here and simply
used the same thermodynamic parameters as in Ref.~\cite{Bron02}
despite there are noticeable differences in some particle number ratios
calculated in the two models. These differences
may be related to the different numbers of resonance states
taken into account and uncertainties in the
decay modes of high excited resonances.

\begin{table}
 \caption{\label{tab:table1}
 Particle number ratios near mid-rapidity in central
Au + Au collisions at $\sqrt s_{NN} = 130$~GeV
calculated with the thermodynamic parameters:
$T = 0.168$~GeV,  $\widetilde{\mu}_B = 0.041$~GeV, $\widetilde{\mu}_S
= 0.010$~GeV and  $\widetilde{\mu}_Q = -0.001$~GeV.
 }
 \begin{ruledtabular}
 \begin{tabular}{lccr}  
particle number ratios &       our MC &  statistical model \cite{Bron02} & experiment \\
\tableline
$\pi^-/\pi^+$       &   $0.98$     &  $1.02$     &   $1.00 \pm 0.02$ \cite{35}, $0.99 \pm 0.02$ \cite{36}   \\
$\bar p/\pi^-$      &   $0.06$     &  $0.09$     &   $0.08 \pm 0.01$ \cite{37}         \\
$K^-/K^+$           &   $0.90$     &  $0.92$     &   $0.91 \pm 0.09$ \cite{35}, $0.93 \pm 0.07$ \cite{39}\\
$K^-/\pi^-$         &   $0.22$     &  $0.16$     &   $0.15 \pm 0.02$ \cite{40}  \\
$\bar p/p$          &   $0.61$     &  $0.65$     &   $0.60 \pm 0.07$ \cite{35}, $0.64 \pm 0.08$ \cite{39}\\
$\bar \Lambda/\Lambda$ & $0.69$    &   $0.69$    &    $0.71 \pm 0.04$ \cite{38}\\
$\bar \Xi/\Xi$       &  $0.79$     &  $0.77$     &   $0.83 \pm 0.06$ \cite{38} \\
$\phi/K^-$           &  $0.17$     &  $0.15$     &   $0.13 \pm 0.03$ \cite{41}  \\
$\Lambda/p$          &  $0.48$     &  $0.47$     &   $0.49 \pm 0.03$ \cite{42}, \cite{43}   \\
$\Xi^-/\pi^-$        &  $0.0086$   &  $0.0072$   &   $0.0088 \pm 0.0020$ \cite{45}  \\

 \end{tabular}
 \end{ruledtabular}
 \end{table}

\begin{table}
 \caption{\label{tab:table2}
Particle number ratios near mid-rapidity in central
Au + Au collisions at $\sqrt s_{NN} = 200$~GeV
calculated with the thermodynamic parameters:
$T = 0.165$~GeV, $\widetilde{\mu}_B = 0.028$~GeV, $\widetilde{\mu}_S = 0.07$~GeV,
and $\widetilde{\mu}_Q = -0.001$~GeV.
 }
 \begin{ruledtabular}
 \begin{tabular}{ccc}
particle number ratios      & our MC   & experiment \cite{pt_PHENIX} \\
\tableline
$\pi^-/\pi^+$          &$0.98$      &  $0.984 \pm 0.004$ \\
$K^-/K^+$              &$0.94$      &  $0.933 \pm 0.008$ \\
$K^-/\pi^-$            &$0.21$      &  $0.162 \pm 0.001$ \\
$\bar p/p$             &$0.71$      &  $0.731 \pm 0.011$  \\
\end{tabular}
 \end{ruledtabular}
 \end{table}

 \subsection{\label{sec92} Pseudo-rapidity distributions}
\begin{figure}
\begin{center}
\begin{tabular}{cc}
\includegraphics[width=8.3cm]{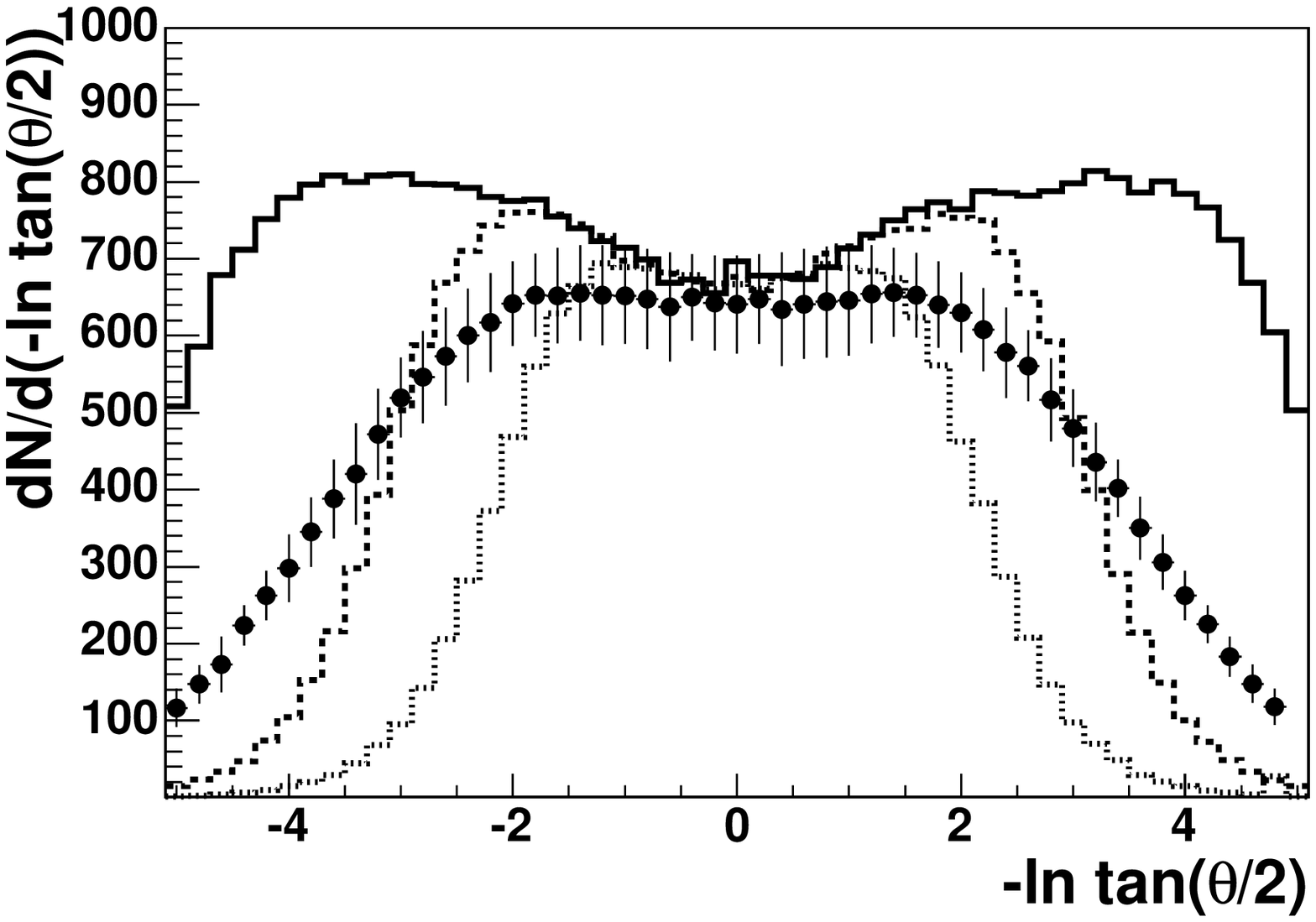}
\includegraphics[width=8.3cm]{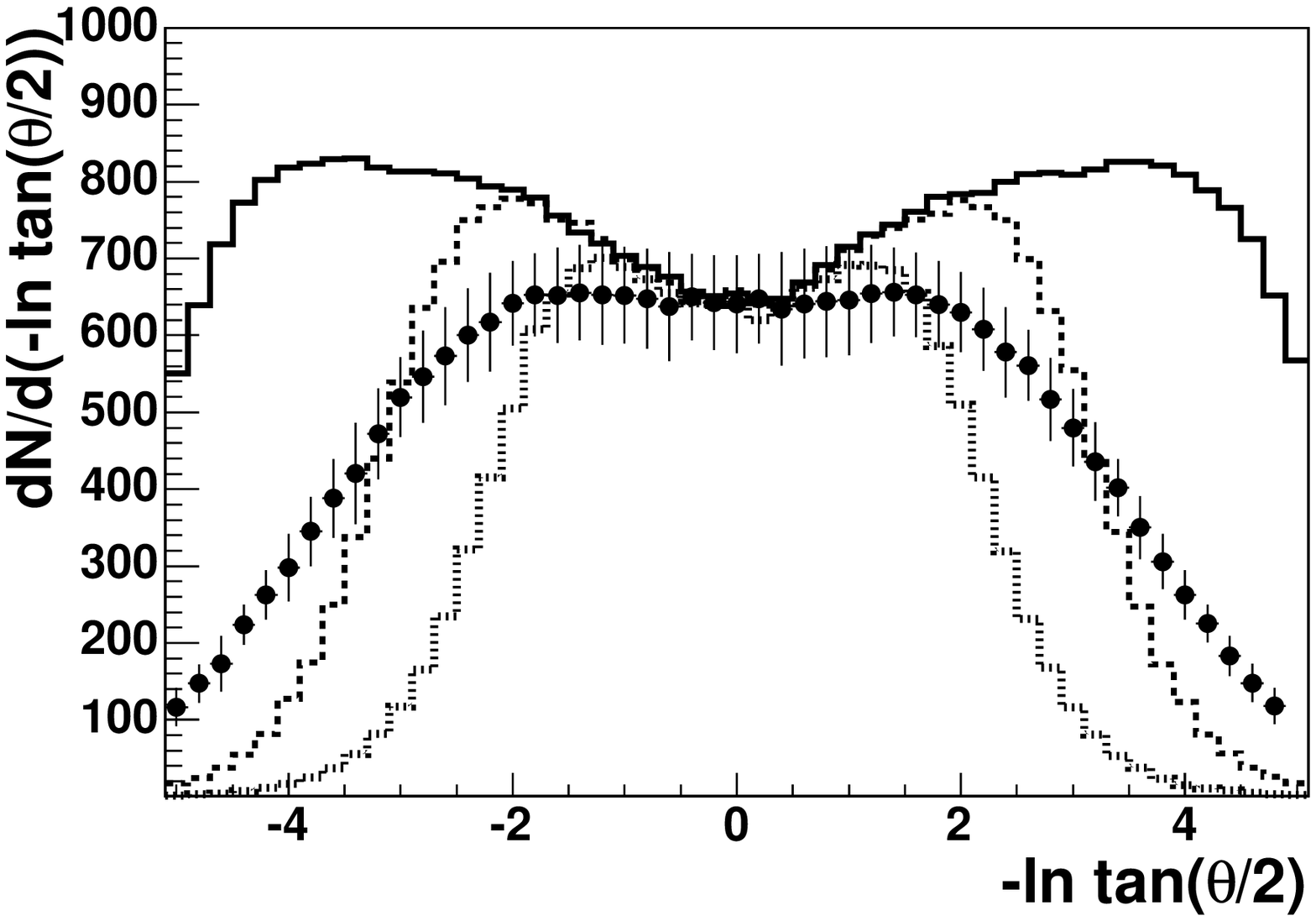}
\end{tabular}
\end{center}
\caption{The pseudo-rapidity
($-\ln\tan(\theta/2)$, $\theta$ is the particle production angle)
distributions of charged particles
in central Au + Au collisions at $\sqrt s_{NN} = 200$~GeV
from the PHOBOS experiment \cite{eta_PHOBOS} (solid circles)
and the MC calculations within the Bjorken-like (left panel)
and Hubble-like (right panel) models. The model results
corresponding to the space-time rapidity range parameter
$\eta_{\max}=5, 3$ and 2 are shown
by solid, dashed and dotted lines respectively.
\label{fig:eta}
}
\end{figure}

In Fig.~\ref{fig:eta}, we compare the PHOBOS data \cite{eta_PHOBOS}
on pseudo-rapidity spectrum of charged hadrons in central Au+Au
collisions at $\sqrt s_{NN}=200$~GeV
with our MC results obtained within the Bjorken-like and Hubble-like
models for different values of $\eta_{\max}$.
One may see that the data are compatible with the longitudinal
boost invariance only in the mid-rapidity region in which
the model is practically insensitive to $\eta_{\max}$.
In the single freeze-out scenario,
the data on particle numbers at mid-rapidity thus allows one
to fix the effective volume $V_{\rm eff}\propto \tau R^2$.

\subsection{\label{sec93} Transverse momentum spectra}
\begin{figure}
\begin{center}
\begin{tabular}{cc}
\includegraphics[width=8.3cm]{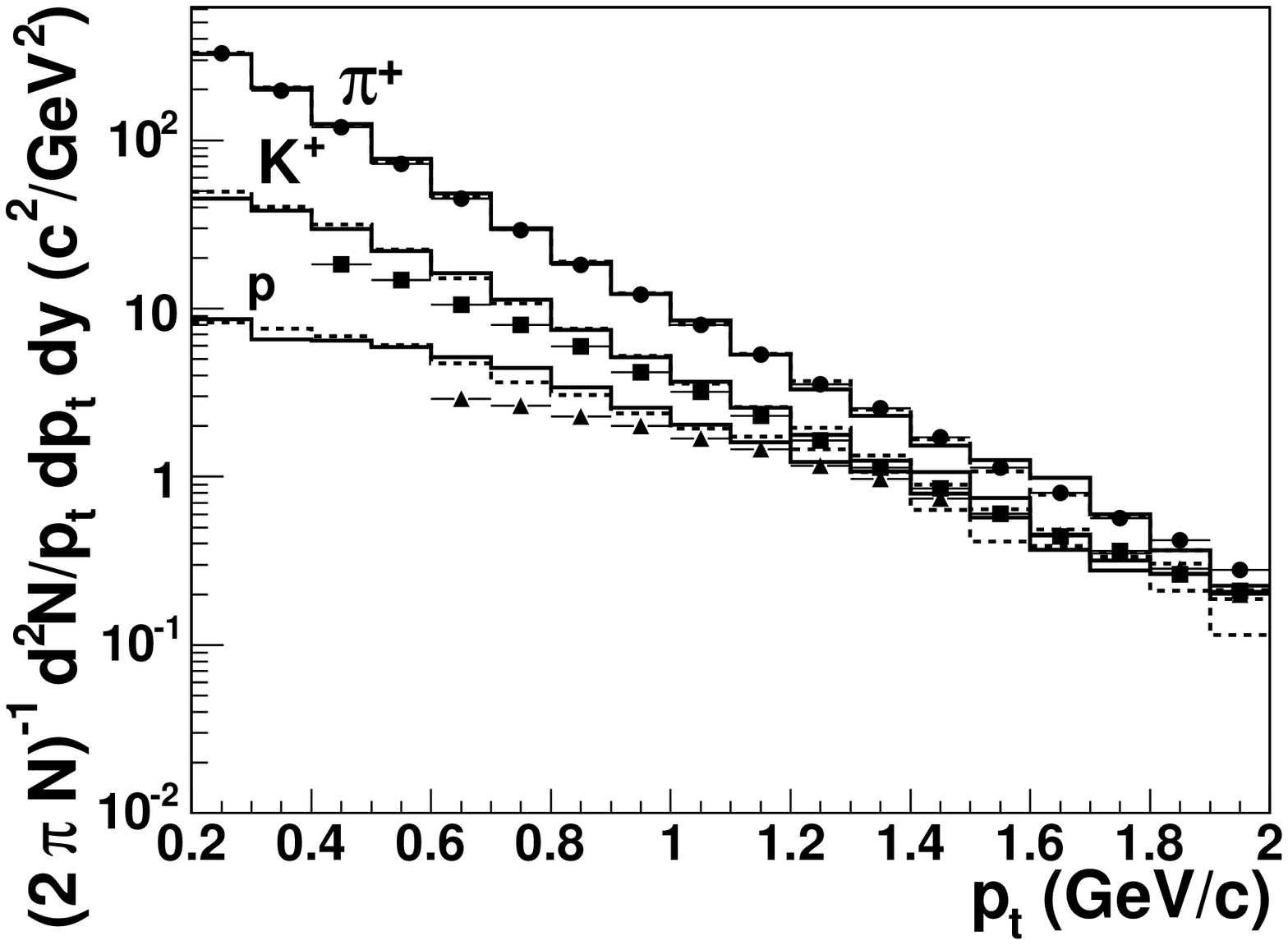}
\includegraphics[width=8.3cm]{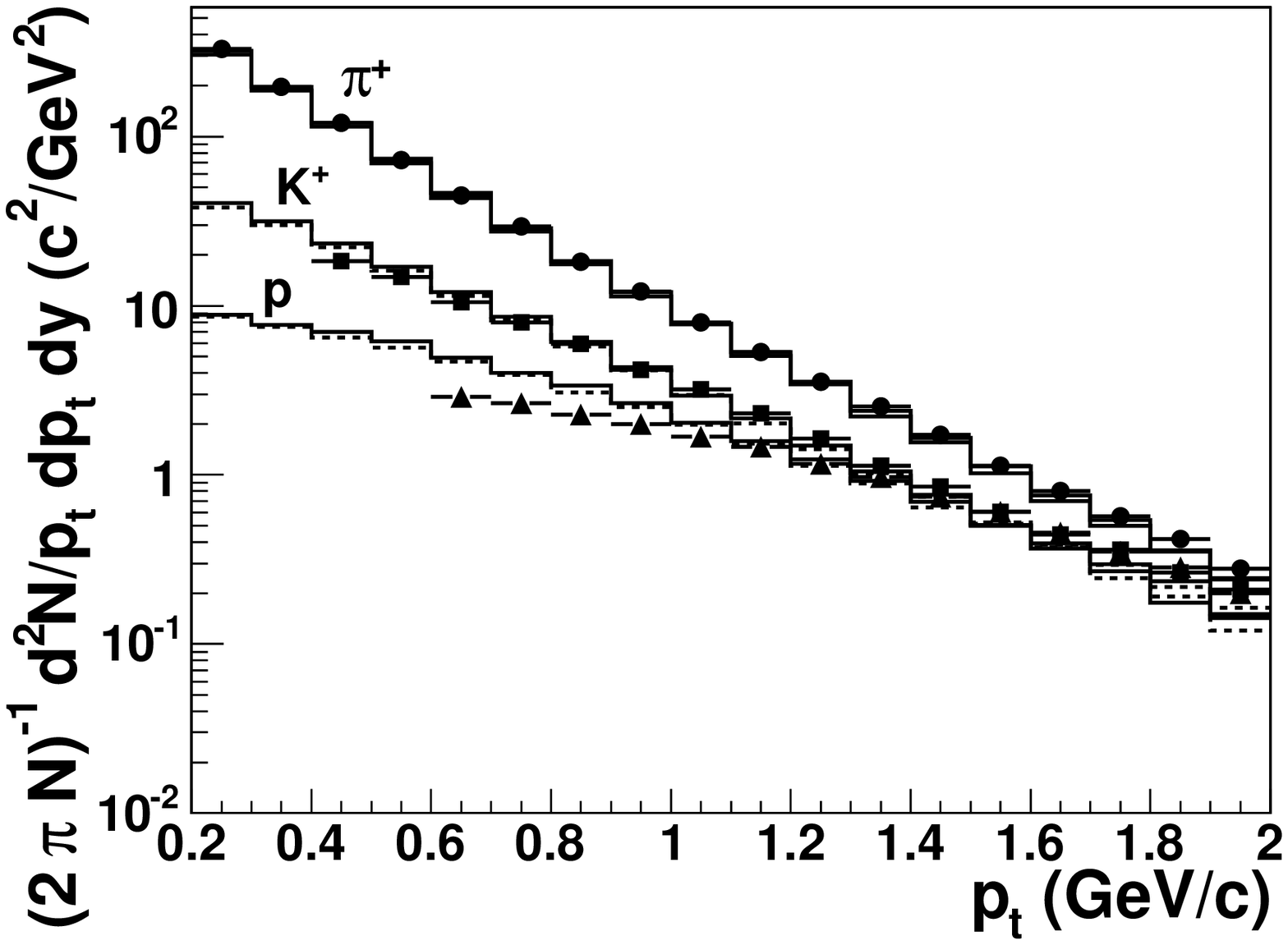}
\end{tabular}
\end{center}
\caption{The $\pi^{+}$, $K^{+}$ and proton transverse momentum spectra
at mid-rapidity $y\approx 0$ in central Au + Au collisions at $\sqrt s_{NN} = 200$~GeV
from PHENIX experiment \cite{pt_PHENIX} (solid symbols)
and the MC calculations within the
Bjorken-like (dashed lines) and
Hubble-like (solid lines) models.
The right panel shows the model results obtained
with the strangeness suppression parameter $\gamma_s=0.8$.
\label{fig:pt}
}
\end{figure}

In Fig.~\ref{fig:pt}, we compare the mid-rapidity
PHENIX data \cite{pt_PHENIX}
on $\pi^+$, $K^+$ and proton $p_t$ spectra in Au+Au
collisions at $\sqrt s_{NN}=200$~GeV
with our MC results obtained within the Bjorken-like
and Hubble-like models.
A good agreement between the models and the data may be seen for
pions while for kaons and protons the models overestimate the
spectra at $p_t < 1$ GeV/$c$. For kaons, this discrepancy can be
diminished with the help of the strangeness suppression
parameter $\gamma_s$ of 0.8 (see the right panel in
Fig.~\ref{fig:pt}).
The overestimated slope of the kaon and proton $p_t$ spectra
can also be related with the oversimplified assumption of a common
thermal and chemical freeze-out or insufficient number of the
accounted heavy resonance states.

\begin{figure}
\includegraphics[width=14.00cm]{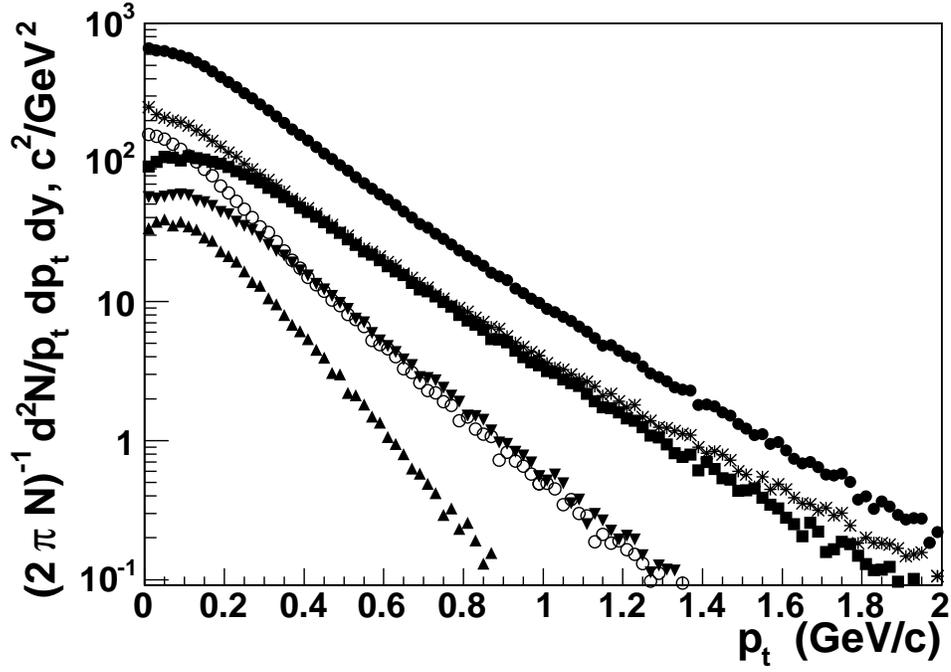}
\caption{The contributions to the $\pi^+$ transverse momentum spectrum at mid-rapidity
in central Au + Au collisions at $\sqrt s_{NN} = 200$~GeV
calculated within the Bjorken-like model:
all $\pi^+$'s (solid circles), direct $\pi^+$'s (stars), decay $\pi^+$'s
from $\rho$ (squares), $\omega$ (open circles), $K^*(892)$ (up-triangles)
and $\Delta$ (down-triangles).
\label{fig:ptspec}
}
\end{figure}

The contribution of different resonances to the pion $p_t$ spectrum
calculated in the Bjorken-like model is shown in Fig.~\ref{fig:ptspec}.

Note that in Hubble-like model, the transverse flow is determined
by the volume parameters $R, \tau$ and so, at fixed
thermodynamic parameters and the effective volume
$V_{\rm eff}\propto \tau R^2$, the transverse momentum
spectra allow one to fix both $R$ and $\tau$.
In the Bjorken-like model, there is more freedom since the
transverse flow is also regulated by the parameter
$\rho_u^{\max}$. The choice of these parameters in Table \ref{tab:table4}
has been done to minimize the discrepancy of the simulated and
measured correlation radii of identical pions (see below).

\subsection{\label{sec95} Correlation functions}
It is well known that, due to the effects of quantum statistics (QS) and final
state interaction (FSI),
the momentum correlations of two or more particles at small
relative momenta in their center-of-mass system
are sensitive to the space-time characteristics of the production
process on a level of fm $=10^{-15}$ m
so serving as a correlation femtoscopy tool
(see, for example, \cite{Lednicky05}-\cite{CF_STAR}).

The momentum correlations are usually studied with the help of
correlation functions of two or more particles. Particularly, the
two-particle correlation function $CF(p_{1},p_{2})$ is defined as a
ratio of the measured two-particle distribution to the reference one
which is usually constructed by mixing the particles from different
events of a given class, normalizing the correlation function to
unity at sufficiently large relative momenta.

Since our MC generator provides the information on
particle four-momenta $p_i$ and four-coordinates $x_i$
of the emission points, it can be used to calculate the correlation
function with the help of the weight procedure, assigning a weight
to a given particle combination accounting for the effects of QS and
FSI. Here we will consider the correlation function of two identical
pions neglecting their FSI, so the weight
\begin{equation}
      w = 1 + \cos(q \cdot \Delta x),
\label{cf4}
\end{equation}
where  $q = p_1 -p_2 $ and $\Delta x = x_1 - x_2$. The $CF$ is
defined as a ratio of the weighted histogram of the pair kinematic
variables to the unweighted one.

Generally, the pair is characterized by six kinematic variables. In
case of the azimuthal symmetry, there are five variables that are
usually chosen as the three "out-side-long" components of the
relative three-momentum vector \cite{pod83, Pratt84}
$\mathbf{q}=(q_\mathrm{out},q_\mathrm{side},q_\mathrm{long})$, half
the pair transverse momentum $k_t$ and the pair rapidity or
pseudo-rapidity.
The {out} and {side} denote the transverse, with respect to the
reaction axis, components of the vector ${\bf q}$; the {out}
direction is parallel to the transverse component of the pair
three--momentum.

The corresponding correlation widths are usually parameterized in
terms of the Gaussian correlation radii $R_i$,
\begin{equation}
CF(p_{1},p_{2})= 1+\lambda\exp(-R_\mathrm{out}^2q_\mathrm{out}^2
-R_\mathrm{side}^2q_\mathrm{side}^2
-R_\mathrm{long}^2q_\mathrm{long}^2
-2R_\mathrm{out,long}^2q_\mathrm{out}q_\mathrm{long}) \label{cf3}
\end{equation}
and their dependence on pair rapidity and transverse momentum is
studied. The form of Eq.~(\ref{cf3}) assumes azimuthal symmetry of
the production process \cite{pod83}. Generally, e.g., in case
of the correlation analysis with respect to the reaction plane, all
three cross terms $q_iq_j$ contribute \cite{Wiedemann_Heinz02}. We
choose as the reference frame the longitudinal co-moving system
(LCMS) \cite{cso91}. In LCMS each pair is emitted transverse to the
reaction axis so that the pair rapidity vanishes. The parameter
$\lambda$ measures the correlation strength. For fully chaotic
Gaussian source $\lambda=1$. Experimentally observed values of
$\lambda< 1$ are mainly due to contribution of very long--lived
sources ($\eta$, $\eta'$, $K^0_s$, $\Lambda$, \dots), the
non-Gaussian shape of the correlation functions and particle
misidentification.

\begin{figure}
\includegraphics[width=14.00cm]{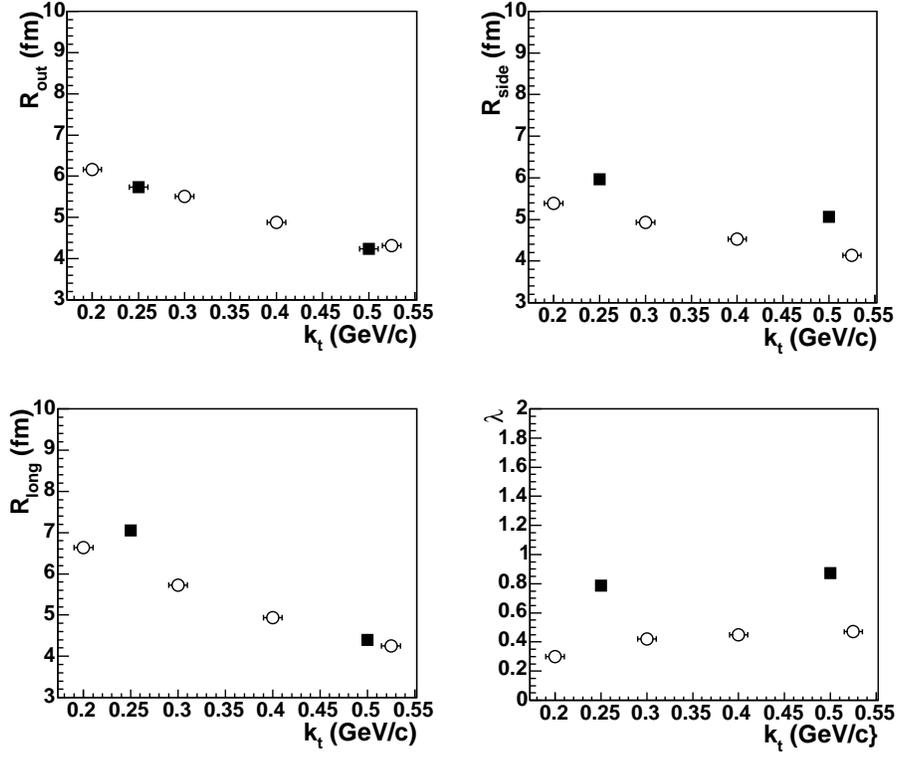}
\caption{The $\pi^\pm \pi^\pm$ correlation radii
and the suppression parameter $\lambda$
at mid-rapidity in central $Au+Au$
collisions at $\sqrt s_{NN} = 200$~GeV from the STAR experiment
\cite{CF_STAR} (open circles) and the MC calculations
within the Bjorken-like model
(up-triangles) in different intervals of the pair
transverse momentum $k_t$.
\label{fig:CF}
}
\end{figure}

The correlation functions of two identical charged pions have been
calculated within the Bjorken-like and Hubble-like models
with the parameters given in Table \ref{tab:table4}, $\eta_{\max}=2$,
reasonably well describing single particle spectra in the
mid-rapidity region. The three-dimensional
correlation functions were fitted according to Eq.~(\ref{cf3}) in
two $k_t$ intervals $0.1<k_t<0.3$~GeV/$c$ and $ 0.3<k_t<0.6$~GeV/$c$. In
Fig.~\ref{fig:CF}, the fitted correlation radii and strength parameter are
compared with those measured by STAR collaboration \cite{CF_STAR}.
One may see that the Bjorken-like model, adjusted to describe
single particle spectra, describes also the decrease of the correlation radii
with increasing $k_t$ but overestimates their values.
The situation is even worth with the Hubble-like model which is
more constraint than the Bjorken-like one and yields
the longitudinal radius by a factor two larger.

As for the overestimation of the correlation strength parameter
$\lambda$, it is likely related to the neglected contribution
of misidentified particles and pions from weak decays.
Indeed, the new
preliminary analysis of the STAR data with the improved particle
identification \cite{Bistr} yields the $\lambda$ parameter closer to
the model results.

We would like to emphasize that the high freeze-out
temperature of $165$~MeV and a fixed effective volume
$V_{\rm eff}\propto \tau R^2$ make it quite difficult
to describe the correlation radii within the single
freeze-out model.
Thus a tuning of the longitudinal radius
$R_{\rm long}\approx \tau(T/m_t)^{1/2}$ requires a small
proper time $\tau$, leading to too large values of $R$ and
$R_{\rm side}\propto R$.
The concept of
a later thermal freeze-out occurring at a smaller temperature
$T^{\rm th}<T^{\rm ch}$ and with no multiplicity constraint on the
thermal effective volume
(see section \ref{sec2}) can help to resolve this problem
(see, e.g., \cite{Lisa}).

To get a valuable information from the correlation data,
one should consider more realistic models as compared with the simple
Bjorken-like and Hubble-like ones
(particularly, consider a more complex form
of the freeze-out hypersurface taking into account particle emission
from the surface of expanding system \cite{Borysova}) and
study the problem of particle
rescattering and resonance excitation after the
chemical and/or thermal freeze-out (only
minor effect of elastic rescatterings on particle spectra and
correlations is expected \cite{UKM05}).
For the latter, our earlier
developed C++ kinetic code \cite{UKM05} can be coupled
to the MC freeze-out generator.

\maketitle
\section{\label{sec10} Conclusions and perspectives}
We have developed a MC simulation procedure and the corresponding
C++ code allowing for a fast but
realistic description of multiple hadron production in
central relativistic heavy ion collisions.
A high generation speed and an easy control through input
parameters make our MC generator code particularly useful
for detector studies.
As options, we have implemented two freeze-out scenarios with
coinciding and with different chemical and thermal freeze-outs.
Also implemented are various options of the freeze-out hypersurface
parameterizations including those with non-space-like
hypersurface sectors related
to the emission from the surface of expanding system.
The generator code is quite flexible and allows the user to add
other scenarios and freeze-out surface parameterizations as well as
additional hadron species in a simple manner.

We have compared the RHIC experimental data with our MC
generation results obtained within the single freeze-out scenario and
Bjorken-like and Hubble-like freeze-out surface parameterizations.
While simplified, such a scenario nevertheless
allows for a reasonable description of particle spectra.
It however fails to describe the correlation functions of
identical pions, overestimating the correlation radii.

The RHIC data thus points to the need for a more complicated
scenario likely including different chemical and thermal
freeze-outs, a more complex form of the freeze-out hypersurface
(the use of numerical solution of the relativistic hydrodynamics
is foreseen) and the account for kinetic evolution following
the chemical and/or thermal freeze-out (for this, the MC generator
can be coupled to our C++ kinetic code \cite{UKM05}).

We plan to implement in the MC generator the impact parameter
dependence of the freeze-out hypersurface and account for
the anisotropic flow similar to Ref. \cite{Lokhtin03, Lokhtin05}.
In view of the importance of high-$p_t$ physics related
to the partonic states created in ultra-relativistic heavy
ion collisions, we also foresee the inclusion of mini-jet production
\cite{Lokhtin05}.

\begin{acknowledgments}
We would like to thank B.V.~Batyunia and L.I.~Sarycheva for useful discussions.
The research has been carried out within the scope of the ERG
(GDRE): Heavy ions at ultra-relativistic energies -– a European
Research Group comprising IN2P3/CNRS, Ecole des Mines de Nantes,
Universite de Nantes, Warsaw University of Technology, JINR Dubna,
ITEP Moscow and Bogolyubov Institute for Theoretical Physics NAS of
Ukraine. This work has been supported, in part, by
Grant of Russian Agency for Science and Innovations
under contract 02.434.11.7074 (2005-RI-12.0/004/022),
by the special program of the Ministry of
Science and Education of the Russian Federation,grant
RNP.2.1.1.5409, by the
Grant Agency of the Czech Republic under contract 202/04/0793
and by 
Award No. UKP1-2613-KV-04 of the U.S. Civilian 
Research and Development Foundation 
(CRDF) and Fundamental Research State Fund of Ukraine, Agreement No. 
F7/209-2004.

\end{acknowledgments}

\clearpage

\clearpage
\newpage

\end{document}